\documentclass[a4paper,fleqn, usenatbib]{mnras}

\usepackage[T1]{fontenc}
\usepackage{ae,aecompl}

\usepackage{graphicx}	
\usepackage{savesym}
\usepackage{amsmath}
\savesymbol{iint}
\usepackage{txfonts}
\restoresymbol{TXF}{iint}
\usepackage{amssymb}

\newcommand{\mr}{\mathrm}
\renewcommand{\vec}[1]{\boldsymbol{\mathbf{#1}}}
\usepackage{mathrsfs}
\usepackage{bm}
\usepackage{url}
\usepackage{float}
\usepackage{subfig}
\graphicspath{{figs/}}

\title{Efficiency of Planetesimal Ablation in Giant Planetary Envelopes} 

\author[A. Pinhas et al.]{
Arazi Pinhas,$^{1}$\thanks{E-mail: ap817@cam.ac.uk}
Nikku Madhusudhan,$^{1}$
Cathie Clarke$^{1}$
\\
$^{1}$Institute of Astronomy, University of Cambridge, Madingley Road, CB3 0HA
}

\date{Accepted XXX. Received YYY; in original form ZZZ}

\pubyear{2016}

\begin{document}
\label{firstpage}
\pagerange{\pageref{firstpage}--\pageref{lastpage}}
\maketitle
\begin{abstract}
Observations of exoplanetary spectra are leading to unprecedented constraints on their atmospheric elemental abundances, particularly O/H, C/H, and C/O ratios. Recent studies suggest that elemental ratios could provide important constraints on formation and migration mechanisms of giant exoplanets. A fundamental assumption in such studies is that the chemical composition of the planetary envelope represents the sum-total of compositions of the accreted gas and solids during the formation history of the planet. We investigate the efficiency with which accreted planetesimals ablate in a giant planetary envelope thereby contributing to its composition rather than sinking to the core. From considerations of aerodynamic drag causing `frictional ablation' and the envelope temperature structure causing `thermal ablation', we compute mass ablations for impacting planetesimals of radii 30 m to 1 km for different compositions (ice to iron) and a wide range of velocities and impact angles, assuming spherical symmetry. Icy impactors are fully ablated in the outer envelope for a wide range of parameters. Even for Fe impactors substantial ablation occurs in the envelope for a wide range of sizes and velocities. For example, iron impactors of sizes below $\sim$0.5 km and velocities above $\sim$30 km/s are found to ablate by $\sim$ 60-80\% within the outer envelope at pressures below $10^3$ bar due to frictional ablation alone. For deeper pressures ($\sim$10$^7$ bar), substantial ablation happens over a wider range of parameters. Therefore, our exploratory study suggests that atmospheric abundances of volatile elements in giant planets reflect their accretion history during formation. 
\end{abstract}



\section{Introduction}
Chemical characterization of exoplanetary atmospheres is the new frontier of exoplanetary science. Numerous observational efforts in recent years have led to conclusive detections of atomic and molecular species, and the estimation of their abundances, in a wide range of exoplanetary atmospheres \citep[see e.g. reviews by][]{madhu14a,madhu16}. The abundances of prominent O and C bearing molecules, such as H$_2$O, CO, and/or CH$_4$ have, in turn, been used to estimate elemental abundance ratios such as O/H, C/H, and C/O ratios where possible \citep[e.g.,][] {madhu11,madhu14c,line12,line14,benneke15,kreidberg15,todorov16}. The derived elemental abundances reveal a diversity of chemical compositions ranging from sub-solar to super-solar metallicities and C/O ratios. However, estimating elemental abundances of exoplanetary atmospheres is a challenging and evolving area and substantial work is underway to derive accurate and precise abundances taking into account the various uncertainties owing to limited spectral coverage of existing instruments and the possibility of diverse clouds/hazes in the atmospheres \citep{sing16}. Nevertheless, with increasing atmospheric observations using large facilities (e.g. HST, VLT, etc.) and increasing numbers of exoplanets found around bright stars, abundances estimates of exoplanetary atmospheres will continue to improve greatly. The prospects are even higher with upcoming large facilities such as JWST and E-ELT. Armed with these facilities, the diverse range of giant exoplanets with temperatures of $\sim$400-3000 K provide unique laboratories to study their chemical compositions and diversity \citep{madhu12}. 

Latest studies are now exploring the exciting possibility of using elemental abundances of giant exoplanetary atmospheres to constrain the conditions and processes of planetary formation and migration. Such an approach has a precedent in the solar system. In Jupiter's atmosphere, the abundances of C, N, S, Ar, Kr, and Xe relative to H have been measured to be $\sim$2-4 times the solar values \citep{owen99,atreya05} which suggest substantial accretion of solids during its formation. These abundances have been used as evidence for Jupiter's metal-rich interior and its formation by core accretion. Core accretion models predict a super-solar abundance (3-7 times) of O in Jupiter's atmosphere \citep{mousis12}. Motivated by early abundance ratio estimates for hot Jupiters \citep{madhu12}, initial studies based on formation models of Jupiter suggested that giant planets that formed via core-accretion in the $\sim$5-20 AU region of a solar-composition disk possessed super-solar metallicities and nearly solar C/O ratios of $\sim$0.5 \citep{mousis09,madhu11}. However, \citet{oberg11} suggested that at large orbital separations the gas in the disk, and hence the forming giant planets, can have high C/O ratios ($\sim$1) as the oxygen-rich volatiles all condense out. However, the compositions of gas and solids accreted depend on the composition and thermodynamic properties of the disk at the given location which are time-dependent \citep{helling14,marboeuf14}. In addition, giant planets can also migrate through the disk \citep{nelson00,papaloizou07}, thereby accreting matter from different regions of the disk which then contribute to the final planetary composition. 
Atmospheric metallicities of hot Jupiters could also potentially constrain their migration mechanisms as migration through the disk is more likely to cause metal enrichment due to solid accretion compared to disk-free migration mechanisms \citep{madhu14b}.

The fundamental assumption in trying to constrain planetary formation from observed atmospheric abundances is that the observed elemental abundances represent the sum-total of elemental abundances in the gas and solids accreted during the formation history of the planet. This assumption is critically important as the metallicity of an atmosphere (e.g. O/H or C/H ratios) is directly related to the amount of accreted solids dissolved in the atmosphere. For example, super-solar metallicities are strong signatures of substantial planetesimal accretion. On the other hand, when sub-solar metallicities are observed it is unclear if they are due to lack of substantial planetesimal accretion or if the accreted planetesimals have not ablated significantly in the atmosphere. Therefore, a primary question to be answered currently is ``to what extent" do accreted planetesimals ablate in the envelope as opposed to sinking to the core partially or fully with limited contribution to the envelope. The present work is focused on addressing this question. 

It can be assumed that the observed atmospheric composition of a giant planet is representative of the composition of the envelope. Convection is believed to be confined to localised (short-scale) zones when the envelope mass is much less than the core mass during protoplanetary evolution, but subsequently gas-giant planets are expected to have envelopes unstable against full convection \citep{pollack86}. It is expected that chemistry circulation takes place on timescales much shorter than formation times so that atmospheric compositions of exoplanetary Jovian analogue systems should reflect the time-averaged input of chemical constituents into the envelope during the late evolutionary and equilibrated phases. 

A plethora of studies exist in the literature on dust, cometary, asteroidal, and planetesimal interactions with protoplanetary, planetary,  and  compact  object  atmospheres. The case of cometary fragments impacting Jupiter, e.g. the 1994 Shoemaker-Levy 9 (SL9) event \citep{pond12,korycansky06,crawford95,maclow94}, allows for studies of impactors in the context of our solar system. The SL9 event concerned the demise of a gravitationally-focused long-period comet which fragmented onto Jupiter's atmosphere during a time span of six days in July, 1994, in which approximately 21 cometary fragments impacted Jupiter. The cometary fragments reached impact velocities of about $60\, \mr{km\,s^{-1}}$ at an angle of $43$ degrees from the Jovian rotation axis \citep{harrington04}. In today's solar system similar events are observed more commonly than traditionally expected \cite[cf.][]{hueso13,pond12} and are believed to have occurred habitually in our natal solar nebula. The discovery of atypical molecules in the Jovian atmosphere implied chemical enrichment of its atmosphere from the event. SL9 event model simulations \citep{pond12,korycansky06,crawford95,maclow94} used a sophisticated 3D hydrodynamical code, ZEUS-MP2, in effort to better understand impactor characteristics through consideration of plume ejecta and shock physics. Although such events are now understood to be slightly more common than hitherto thought, present-day accretion of solids is unlikely to significantly alter the heavy-element content of a 4.5 Gyr Jupiter. Though our work uses the structure of a cold, present Jupiter, we ultimately aim to provide a benchmark study given the many works conducted on SL9.

Others have studied occurrences on Earth such as the Chelyabinsk \citep{korycansky15,korycansky14} and the the June 30, 1908 Tunguska event \citep{chyba93} in an effort to infer properties of the ablated impactors commensurate with available data. The latter modeled impactors with aerodynamic drag and gravity whilst also accounting for ablation, varying impactor angles and compositions. They principally found that iron impactors penetrate the surface whilst carbonaceous and other materials of low density ablate in the atmosphere. Furthest  from  home,  the  pollution  of  white dwarf photospheres from the accretion of many small asteroids has been studied, using photospheres as tools to measure the bulk elemental composition of extrasolar minor planets \citep{jura08}. \citet{mordasini15} studied global planetary population synthesis models which consider the infall of planetesimals into protoplanetary envelopes in a paradigm of migrating protoplanets. They describe in brief how initial conditions (i.e., initial planetesimal radius and  planetary envelope mass) determine whether the impactor reaches the core without total ablation. The works of \citet{pollack86} and \citet{podolak88} have also explored chemical enrichment of core-accretion envelopes in the stages before the core reaches its critical mass value for `runaway' gas accretion. Considering the formation mechanism of gravitational instability, \citet{helled09} model heavy-element enrichment of a Jovian protoplanet for different orbital distances from the host star. 

Our work focuses upon planetesimal (`bolide') accretion into Jovian-like envelopes.\footnote{ We use the terms `planetesimal', `bolide', and `impactor' interchangeably throughout our work.} We investigate how planetesimals ablate and thus chemically enrich Jovian-like exoplanetary envelopes by following the evolution of impactors with different compositions and initial conditions (ICs). Our work inquires more deeply into a broader range of ICs than the above works and offers a study of the `worst-case' (i.e., deepest penetration) scenario of pure iron bolides with initial comparison to other compositions. It can generally serve as a simple window into evolution of impactors of arbitrary composition impacting Jovian-like exoplanets. The compositions, sizes, and material strengths of accreted planetesimals will have different effects on the composition of planetary envelopes. Indeed, atmospheric enrichment values may even have the possibility of discriminating between core-accretion and gravitational instability formation scenarios, with a higher stellar-normalized enrichment favoring the former \citep{mordasini15}.

In this paper, we present a model for planetesimal accretion onto Jovian circumcore layers to determine the degree to which gas-dynamical pressure induces evaporation and environmental heating causes ablation. These depositions will subsequently experience mixing through convective diffusion, transporting material to higher altitudes close to Jovian cloud layers. This paper is organized as follows. The planetary structure of a Jovian analogue is discussed in Section \ref{ps}. This planetary model is then used as a background for the evolution of planetesimal impactors in Section \ref{impactor evolution and dynamics}. The bimodality of ablation is explored in Section \ref{thermal ablation}, in which we consider a simple model for thermal ablation with the pre-calculated frictional ablation of Section \ref{FA}.\footnote{Our use of ablation terminology differs from that of \citet{mordasini15}. Their `thermal ablation' is our frictional ablation. Moreover, their use of `mechanical ablation' refers to their model of mechanical Raleigh-Taylor instabilities that fragment planetesimals into large numbers of offspring that are eventually `thermally ablated' (our frictional ablation). They therefore don't consider the thermal ablation which we model. The reason for this difference in terminology will become clearer in Section \ref{impactor evolution and dynamics}.} We follow this with a sample of results, primarily of iron bolides, in Section \ref{results} and discuss model limitations and comparisons in Section~\ref{discussion}. We then summarize and conclude our work with a review of the essential outcomes of our study in Section \ref{conclusions}, with a view towards the future.

\section{Planetary Structure}\label{ps}
Any interpretation of impacting planetesimals' dynamics presupposes a knowledge of the planetary structure and composition. We herewith detail our fiducial model for the planetary structure of a Jovian-like planet. 

\subsection{Internal Structure Model}
The interior structure of gas giant planets is thought to be demarcated into three or four basic layers. Our work utilizes the three-layer Jovian model of \citet{becker14} to obtain an adiabatic profile (in which approximately no heat is transferred upon rapid expansion and compression) that is consistent with current observable constraints, using the top two layers as backgrounds for planetesimal evolution in Section \ref{impactor evolution and dynamics}. Two adiabatic fluid layers, the outer and inner envelopes, surround a rocky core and the envelopes are assumed to be homogeneously mixed with differential compositions of hydrogen and helium. We do not include metallic contributions to the structure model because of the trace quantities of heavier elements and molecules in Jupiter. This is of course a limitation to reality, but captures the highly dominant compositions. Furthermore, we are not concerned with compositional modelling of the core and treat its surface as a boundary condition at which our structure equations and potential planetesimal evolution cease. The locations of layer boundaries and the composition in the layers are fitted with values that match observational constraints within the errors (see next paragraph). 

Our fiducial EOS for the planetary structure from \citet{becker14} is created through a patchwork of EOS data and simulations for hydrogen and helium (HREOS.3 and HeREOS.3) that spans large domains in density, pressure, and temperature spaces applicable to gas giant planets and brown dwarfs.\footnote{The hydrogen and helium Rostock EOS data are publicly available at \url{http://vizier.cfa.harvard.edu/viz-bin/VizieR?-source=J/ApJS/215/21}.} 
This was achieved principally through performance of extended {\it ab initio} simulations using density functional theory molecular dynamics (DFT-MD) in the high density and pressure regimes, where experiments are few and limited and quantum correlations need to be considered to account for pressure dissociation and ionizations that occur and persist. These DFT-MD computations for sampled $\rho-T$ values are connected to other portions of the parameter space through a summation of other theoretical chemical models, the result of which generally ensures a consistency in thermodynamics to better than 1 percent \citep{becker14}. The Jovian adiabat interior structure of \citet{becker14} is attractive because it reproduces a number of observational constraints. The one-bar temperature, Jovian mass, Jovian radius, atmospheric mass abundance of helium, angular velocity, and the three non-vanishing, smallest-order moments of gravity used in the `theory of figures' ($J_2$, $J_4$, and $J_6$) are all reproduced within the observational errors. Though the structure we use satisfies the gravitational moment and angular velocity criteria, our model still assumes no rotation in practice because we model the structure as spherically symmetric.

\subsection{Planetary Construction}
\begin{figure*}
    \centering
    \includegraphics[scale=0.8]{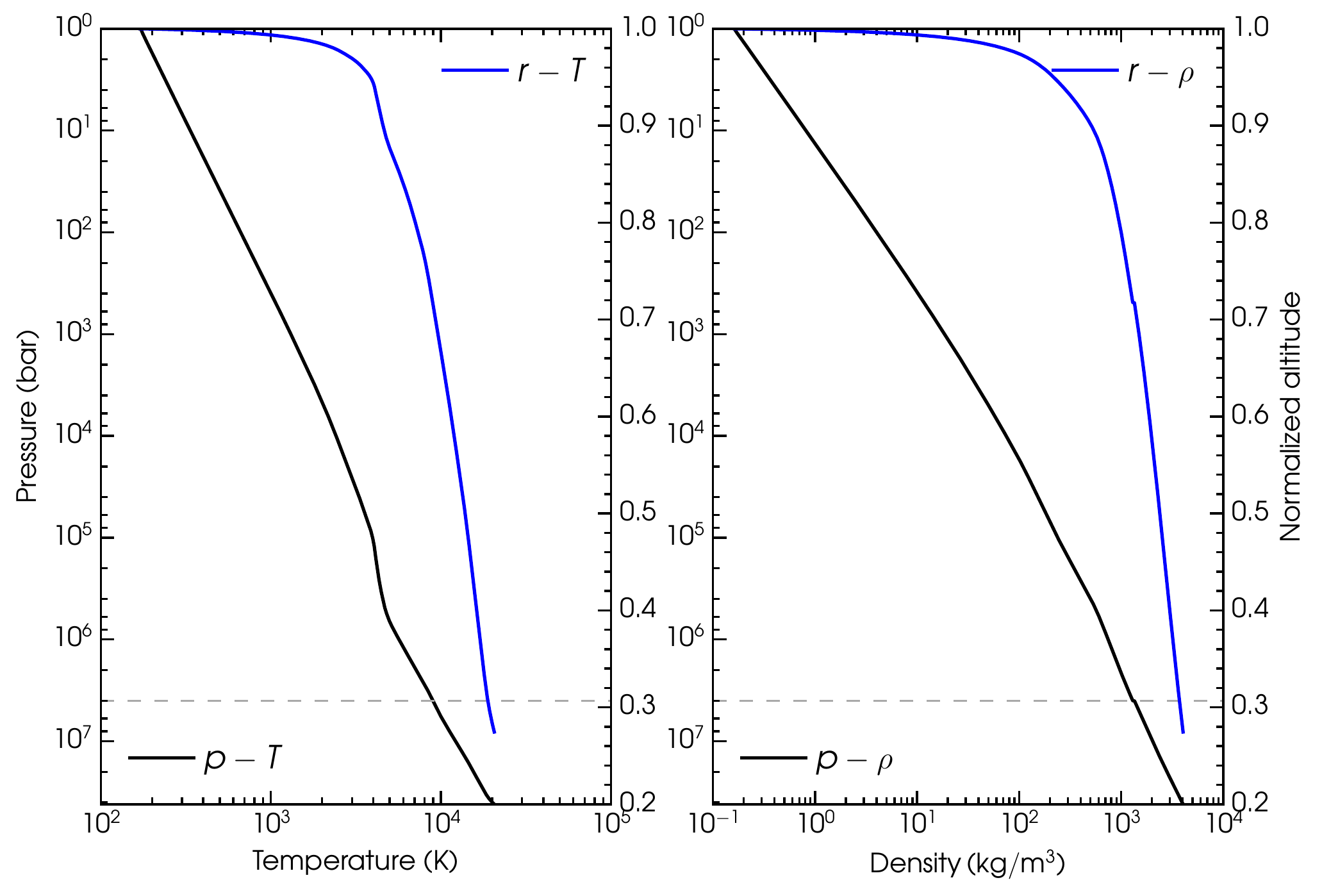}
    \caption{Jovian adiabat structure. Left panel: $p-T$ and $r-T$ profiles shown in black and blue, respectively, based on the linearly-mixed hydrogen and helium Rostock.3 EOSs of \citet{becker14}. The gray dashed isobar shows the insulator-metallic transition at $p=4\,\mr{Mbar}$. The pressure ranges from 1 bar to the core-inner envelope boundary at 41.687 Mbar. Right panel: $p-\rho$ and $r-\rho$ profiles shown in black and blue, respectively, based on the same EOSs and with the same insulator-metallic isobar as in the left panel.}
    \label{fig:planetarystructure}
\end{figure*}
To obtain the three fiducial layers we solve three coupled, ordinary differential equations for the structure of a spherically-equilibriated planet using a fourth-order Runge-Kutta schema: 
\begin{equation}
    \frac{dm}{dp}=-\frac{4\pi r^4}{Gm},
\end{equation}

\begin{equation}
    \frac{dr}{dp}=-\frac{1}{g\rho},
\end{equation}

\begin{equation}
    \frac{\partial T}{ \partial \rho}=\frac{T}{\rho^2}\frac{\left (\frac{\partial p}{\partial T}\right)_{\rho}}{\left(\frac{\partial u}{\partial T}\right)_{\rho}}.
\end{equation}
The combined density is computed via a weighted linear mixture as 
\begin{equation}
    \frac{1}{\rho}=\frac{X}{\rho_\mr{H}}+\frac{Y}{\rho_{\mr{He}}},
\end{equation}
where $X$ and $Y$ are the fractional mass abundances respectively defined as $X=m_\mr{H}/(m_\mr{H}+m_{\mr{He}})$ and $Y=m_{\mr{He}}/(m_\mr{H}+m_{\mr{He}})$. 
Numerical integration requires a unique set of boundary conditions and compositions for hydrogen and helium. We thus set the adiabatic boundary condition to a temperature of $T=170$ K at a pressure of 1 bar. The mass abundances of hydrogen and helium in the outer and inner fiducial envelopes are taken as $X_{\uparrow}=0.762$, $Y_{\uparrow}=0.238$, $X_{\downarrow}=0.709$, and $Y_{\downarrow}=0.291$ \citep{nettelmann12}, where $Y_{\uparrow}$ agrees with Jupiter's atmospheric helium observational constraint and we choose the layer transition at $p_{{\uparrow}-{\downarrow}}=4$ Mbar of the \citet{nettelmann12} J11-4a model. Additionally, the averaged radius at the upper boundary of the outer envelope is $R_J=69911\,\mr{km}$ and we curtail integration at the core-inner envelope boundary at a pressure of $p \equiv 41.687$ Mbar. 

Figure~\ref{fig:planetarystructure} shows the internal structure relations between pressure, altitude, temperature, and density. The linearly-mixed H-He Jovian adiabat extends in temperature from 170 K at the upper boundary to approximately 20kK at the core-inner envelope demarcation of $p = 41.687$ Mbar and $r_{\mr{norm}}=0.275$, and with density flowing over nearly four orders of magnitude from $10^{-1}\,\mr{kg\,m^{-3}}$ to $\sim4 \times 10^{3} \,\mr{kg\,m^{-3}}$. The ideal gas limit extends from the upper 1 bar boundary of the outer envelope to $\sim 10^2\, \mr{kg\,m^{-3}}$, beyond which non-ideal effects such as pressure dissociation and ionization begin to gain influence. The insulator-metallic transition above which most of hydrogen is in molecular form and below which metallic liquid hydrogen becomes physically favorable occurs at 4 Mbar and is shown graphically by the horizontal gray dashed curves for the $p-T$ and $\rho-T$ distributions. 

\section{Impactor Evolution and Dynamics}\label{impactor evolution and dynamics}

We model the dynamics and ablation of planetesimals falling into the structured planet of Section \ref{ps}. \citet{field95} describe several processes able to ablate material from hypersonic compact planetesimal kernels. First, the effect of pressures from material-flows below the planetesimal removes material from the surface when in excess of bolide yield strength; second, ambient thermal heat penetrates the bolide, melting surface material; third, there may arise hydrodynamic instabilities such as Kelvin-Helmholtz and Rayleigh-Taylor (RT) instabilities (e.g., as was believed for SL9); fourth, radiative diffusion due to thermal radiation promotes melting and subsequent atomisation. This list of four processes is by no means exhaustive and a plethora of more detailed considerations in ablating objects may be found in the classic work of \citet{opik58}. The third consideration would induce fragmentation due to aerodynamic stresses, causing for example RT fingers to form along vulnerable sites.

\citet{mordasini15} show by their Figure 9 that fragmentation is minimal for planetesimal radii considered in this work of $R_{pl}\lesssim\mr{1\,km}$. Planetesimals with initial radii of $R_{pl}\gtrsim\mr{1\,km}$ are considerably more sensitive to mechanical instabilities leading to fragmentation. However, planetesimals with radii of $R_{pl}\gtrsim\mr{100\,km}$ do not experience dissociation by mechanical fracture because of their characteristically high self-gravities. We herewith restrict our impacting planetesimal radii from 30 m to 1 km throughout our work. A discussion of analytical models of mechanical instability leading to fragmentation and lateral spread due to differential ram pressures round a bolide is available in \citet{svetsov95}.

In the present work we consider only the first two mechanisms from which we develop a paradigm of mechanical and thermal ablation of impactors interacting with Jovian-analogue envelopes. The third and fourth phenomena would act to place the complete ablation of the impactor proportionally higher in the structured envelopes than when including only the first two mechanisms; our model consequently places lower-limits on the wholesale ablation altitudes which more sophisticated models should only lie above.

\subsection{Frictional Ablation}\label{FA}
\subsubsection{Gravity and Drag}
Infalling planetesimals are accelerated by the gravity of the planet and slowed by aerodynamic drag. The gravitational force between planet and planetesimal and the drag force driving the planetesimal towards terminal velocity are given in the two-body planeto-centric reference frame, by:
\begin{equation}\label{rdotdot}
M_{pl}(\ddot{r}-r\dot{\theta}^2)=-\frac{GmM_{pl}}{r^2}+\frac{1}{2}C_D\rho |\dot{\vec{r}}|_r^2S,
\end{equation}
where $M_{pl}$ is the planetesimal mass; $C_D$ is the drag coefficient; $\rho$ is the ambient density encountered by the planetesimal; $S$ is the cross-sectional area of the planetesimal; and $|\dot{\vec{r}}|_r$ the radial projection of planetesimal velocity. The linear dimensions of the planetesimals are larger than the typical mean free path of the envelope particles (i.e., the Knudson number is $\ll 1$) and the particles thus form a hydrodynamic cushion or condensation cap anterior to the bolides for which thermal velocities may be neglected in light of the relatively high impactor speeds. This is in contrast to scenarios with Knudsen values above unity, for which one must consider the more difficult physics of two-body impacts. The total drag force on the impactor in Equation (\ref{rdotdot}) is hence due to Stokes drag, $\vec{\mathscr{F}}_S=-\frac{1}{2}C_D\rho (\dot{r}^2+r^2\dot{\theta}^2)S$ where the ram pressure exerted on the grain is $\rho |\dot{\vec{r}}|^2$.  The drag coefficient $C_D$ is nominally a function of the velocity of the bolide relative to the ambient medium, the size of the impactor relative to the mean-free-path of envelope molecules, and the fluid viscosity. Indeed, from the Buckingham Pi theorem, one may show that the drag coefficient is generally a pure function of the Reynolds number. For current computational feasibility, however, we can reasonably assume its constancy at unity, consistent with typical estimates; \citet{maclow94} assume unity, \citet{field95} assume 1.2, and \citet{chyba90} value it at 0.92. 

\begin{figure}
\centering
  \includegraphics[scale=.78]{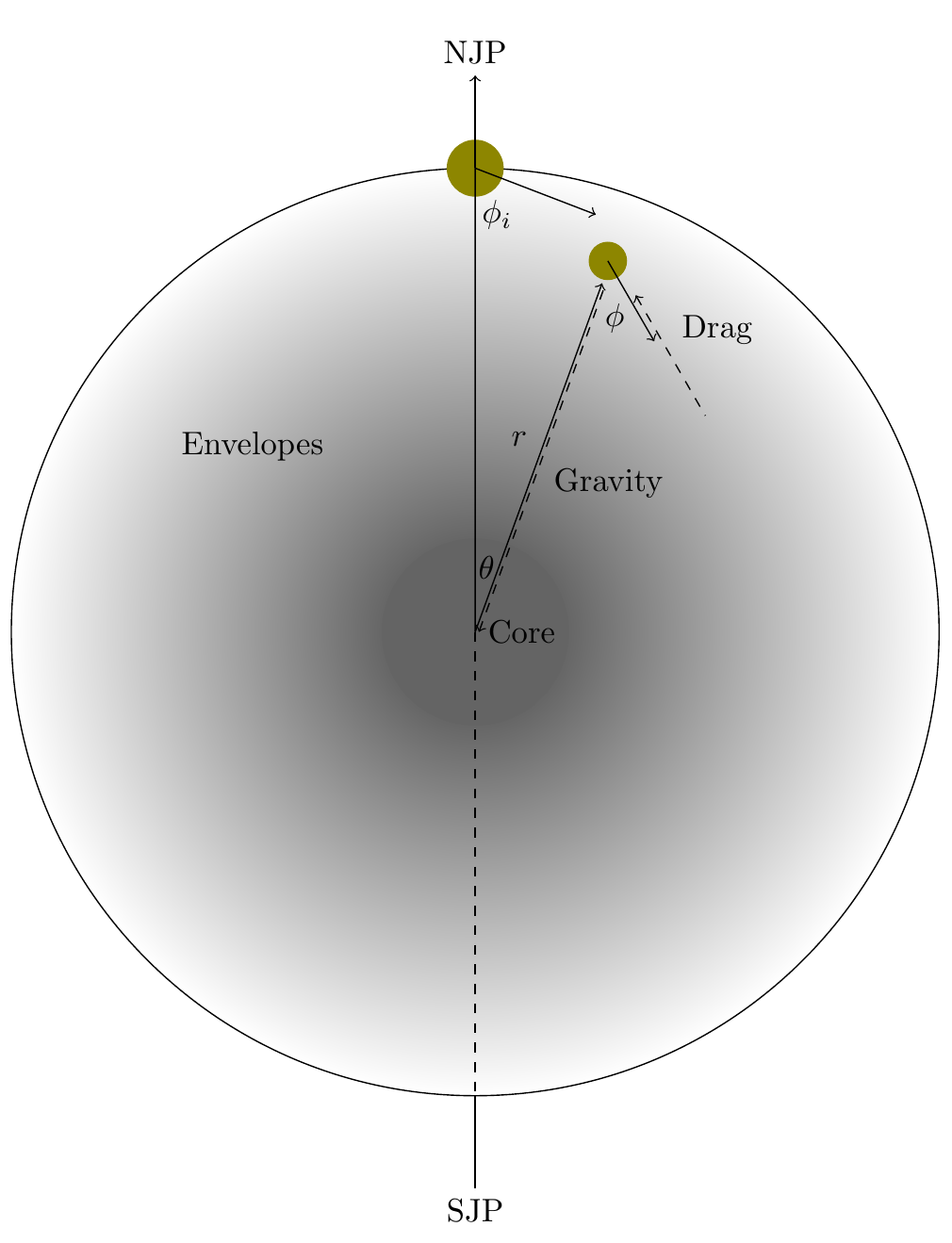}
  \caption{An analogue Jovian planet showing the core and envelopes. The geometry of our fiducial angles is displayed with its assumed local zenith-nadir axis relativity, with the particular zenith and nadir here being the NJP and SJP.}
  \label{geometry}
\end{figure}

Planetesimal influx generally spans all possible initial trajectories relative to local zeniths. We must therefore include an IC of entry angle into our model for a range of non-zero angular  momentum bolides. Figure~\ref{geometry} shows our impactor-planet system geometry. The local zenith chosen in Figure~\ref{geometry} has been taken as the NJP as a specific case, but our results are general to any local zenith. Drag forces always act anti-parallel to the instantaneous impactor trajectory or total velocity vector and the angle $\theta$ is subtended from the local zenith-centre line to the impactor location at time $t$. The angle $\phi$ is the angle between the radial velocity vector and total velocity vector at time $t$ determined instantaneously through $\phi=| \mr{arctan}(r\dot{\theta}/\dot{r}) |$. Regardless of initial impactor angle, we assume all impactors begin from a local zenith in the $r$-$\theta$ plane with projected velocities $v_{r,i}=|\dot{\vec{r}}_i|\cos\phi_i$ and $v_{\theta,i}=|\dot{\vec{r}}_i|\sin\phi_i$. We also take as axiomatic that differential gradients in the drag forces at the planetesimal's sides do not induce a significant torque upon the bolide; induced bolide spins are therefore not modeled. One may question whether this assumption does not lead to major deviations from reality by neglecting what may be an interesting investigation into Magnus force effects because variations in the impactor entry angle have a non-negligible effect on mass ablation loss (see Section \ref{income angle variations}). However, we can carry out a simple calculation to gauge its importance. The lift force coefficient resulting from a spinning planetesimal moving through a medium of density $\rho$ may be written as $C_L=(\vec{F_{mag}} \cdot \hat{\vec{n}})/(1/2S \rho |\dot{\vec{r}}|^2)$ where the unit normal is equal to $\hat{\vec{n}}=\vec{\omega} \wedge \dot{\vec{r}}/|\vec{\omega} \wedge \dot{\vec{r}}|$. \citet{forbes15} shows that when the Reynolds number of our two-body system is sufficiently large ($\gtrsim 10^5$) observational data suggest that $C_L$ nearly approaches the \citet{rubinow61} limit. \citet{rubinow61} show that the Magnus force lift coefficient for a spinning sphere is $C_L=2(R_{pl}\omega/|\dot{\vec{r}}|)|\hat{\vec{\omega}} \wedge \hat{\dot{\vec{r}}}|$. For the most extreme and favorable scenario of spin rotation axis perpendicular to the vector of velocity, i.e. $|\hat{\vec{\omega}} \wedge \hat{\dot{\vec{r}}}| \equiv 1$, and bolide properties that experience highest lift forces (i.e., bolides of $R_{pl}=30\,\mr{m}$, $\omega=2 \times 10^{-2}$ rad\,$\mr{s^{-1}}$ \citep{warner09}, and $|\dot{\vec{r}}|=10 \,\mr{km\,s^{-1}}$), $F_{\mr{mag}}$ is about $10^{-4}\vec{\mathscr{F}}_S$. Larger planetesimals will generally experience smaller lift forces than this. We thus find that our assumption of neglecting both inherent and induced rotation and their associated Magnus forces is substantiated given this small effect and greatly simplifies our work. The cross-section for the aerodynamic force of a spherical bolide is always perpendicular to the vector of loading and is $S(t)=\pi R_{pl}^2(t)$. 
The fiducial geometry allows us to generate our first two equations of motion for the radial and angular components, providing our first two important evolution equations:
\begin{align}
\ddot{r}&=r \dot{\theta}^2-\frac{Gm}{r^2}+\frac{C_D\rho (\dot{r}^2+r^2\dot{\theta}^2) \mr{cos}\phi S}{2M_{pl}},\label{evol1}\\
\ddot{\theta}&=-\frac{2\dot{r}\dot{\theta}}{r}-\frac{C_D\rho (\dot{r}^2+r^2\dot{\theta}^2) \mr{sin}\phi S}{2rM_{pl}}.\label{evol2}
\end{align}

Our inquiry into the evolution of the planetesimal is concerned principally with the extent of vaporisation and dissociation in the envelopes. Following \citet{opik58}, the kinetic energy of envelope swept by a planetesimal in unit time is $S\rho |\dot{\vec{r}}|^3/2$. Of this value, a fraction $C_H$ is utilized in ablating and atomizing the planetesimal and therefore the total rate of mass ablation, and our third evolution equation is, 

\begin{equation}
\frac{dM_{pl}}{dt}=-\frac{C_H \rho |\dot{\vec{r}}|^3S}{2Q_{abl}}\label{mpldot}
\end{equation}
for which $C_H$ is the heat transfer coefficient quantising lended kinetic energy from the anterior particles and $Q_{abl}$ is the specific energy value ({\it i.e.}, per unit mass) necessary in severing the chemical bonds of a kilogram of planetesimal substance and is the net characteristic heat of ablation. The latter is the summation of the proper latent heats of vaporisation and fusion and the intermediate energy in heating the bolides. The bulk kinetic transfer imparted to bolide particulates causes ablation of surface material and we therefore call this kind of ablation `frictional ablation' to distinguish it from the thermal ablation discussed in Section \ref{thermal ablation}. A nominal approach would take $C_H$ and $Q_{abl}$ as functions of impactor morphology, composition, mass, and velocity. Observations have indicated that the heat transfer coefficient remains roughly invariant, and changes only when radiative ablation from thermal radiation by the shocked gas anterior to the impactor becomes dominant over mechanical ablation \citep{bronshten83}. Our model does not include the explicit postshock effects that may yield increased thermal radiation; though a limitation, it allows us to confidently allow for invariance in $C_H$ which we value at 0.01 \citep{chyba90}. 

\subsubsection{Evolution of Planetesimal Radius from Ablation}
The impactor mass $M_{pl}$ varies with altitude as ${dM_{pl}}/{dr}$ and, thus, $S$ likewise evolves as the planetesimal ablates and shrinks. Lateral spread of the impactor, which tends to increase ${dR_{pl}}/{dt}$, is believed to be important for small objects considered in this work, but for simplicity we neglect this inclusion in the present work and focus only on spherical impacting bodies (see Section \ref{discussion} for a more thorough discussion). We henceforth assume homogeneous material ablation from the surface of the bolide and calculate the radial gradient by differentiating the bolide mass $M_{pl}=\frac{4}{3}\pi R_{pl}^3 \rho$ with respect to time to get,
\begin{align}
\frac{dR_{pl}}{dt} &=\frac{R_{pl}}{3 M_{pl}}\frac{dM_{pl}}{dt} \\
\frac{dR_{pl}}{dt} &= -\frac{R_{pl}}{6M_{pl}}C_H\rho |\dot{\vec{r}}|^3 \frac{S}{Q_{abl}} \label{evol5}
\end{align}

\subsubsection{Coupled Solution of Evolution Equations}
The four coupled evolution equations ((\ref{evol1}), (\ref{evol2}), (\ref{mpldot}), (\ref{evol5})) must now be solved numerically. We model planetesimals' interactions with the envelopes using a fourth-order Runge-Kutta finite-difference numerical scheme that integrates the planetesimals' accelerations and frictional ablation. Runge-Kutta requires we generate {\it first order} ordinary differential equations. We thus seek a redefinition of physical parameters. Our redefinition of parameters ${y_i}\, \mr{with}\, i \in[1,6] \cap \mathbb{N}$ are: $y_1\equiv \dot{r}$, $y_2\equiv r$, $y_3\equiv M_{pl}$, $y_4\equiv {R}_{pl}$, $y_5\equiv \dot{\theta}$, $y_6\equiv \theta$. These allow a means to a solution of the four evolution equations alongside two tautological equivalences (for a total of six coupled equations):
\begin{align}
\dot{y}_1&=-\frac{Gm}{y_2^2}+\frac{C_D\rho(y_1^2+(y_2y_5)^2)\cos \phi\, S}{2y_3}+y_2y_5^2 \label{y1}\\
\dot{y}_2&=y_1 \\
\dot{y}_3&=-\frac{1}{2}C_H\rho(y_1^2+(y_2y_5)^2)^{3/2}\frac{S}{Q_{abl}} \label{y3}\\
\dot{y}_4&=-\frac{y_4}{6y_3}C_H\rho(y_1^2+(y_2y_5)^2)^{3/2}\frac{S}{Q_{abl}}\label{y4}\\ \dot{y}_5&=-\frac{C_D\rho(y_1^2+(y_2y_5)^2)\sin \phi S}{2y_3y_2}-\frac{2y_1y_5}{y_2} \label{y5}\\
\dot{y}_6&=y_5 \label{y6}
\end{align}
Equations (\ref{y3}-\ref{y4}) are forced to zero when the average interior aerodynamic force per planetesimal cross-section area is less than the yield strength of the impactor, $p_S/2=\mathscr{F}_S/(2S)<\mr{yield\,\,strength}$, for there is then approximately sufficient structural integrity to arrest the dissociation of the impactor. Again, the magnitudes of the drag force in the radial and azimuthal vectors are determined by the angle carved between $-\hat{\vec{r}}$ and $\vec{v}$ with $\phi=| \mr{arctan}(r\dot{\theta}/\dot{r}) |$. Consistent with average estimates, our work assumes throughout that $C_H=0.01$ and $C_D=1$ from aerodynamic theory of hypervelocity impactors in the terrestrial atmosphere (Melosh 1989).

\subsection{Thermal Ablation}\label{thermal ablation}

We have thus far integrated the physics of frictional ablation in our impactor-planetary model. We now look to a simple model of ambient thermal ablation of an impactor. As an impactor decelerates in the high-density regions, thermal ablation begins operating due to combined effects of the increasing envelope temperatures and thermal conductivities. The melting and subsequent vaporisation is due to pure heating of the bolide from the temperature disparity between planetesimal and ambient envelope. The description of heat transfer within an impactor is modelled using the spherically-symmetric heat diffusion equation, 
\begin{equation}\label{one-d heat equation}
\frac{\partial T}{\partial t} =\frac{\alpha} {r^2}\frac{\partial}{\partial r}\left \{r^2\frac{\partial T}{\partial r} \right\},
\end{equation}
wherein $\alpha \equiv \kappa/(\sigma \rho_i)$ is the thermal diffusivity of the material; $\kappa$ is the thermal conductivity; $\sigma$ is the specific heat capacity for the impactor's composition; and $\rho_i$ is the impactor density. To solve this equation, we proceed with a method similar to that discussed by Recktenwald (2011) which outlines the finite-difference method for obtaining numerical solutions to Equation (\ref{one-d heat equation}).\footnote{See \url{http://www.nada.kth.se/~jjalap/numme/FDheat.pdf}.} Adopting the {\it forward time, centered-space or FTCS approximation} to a finite-difference solution allows us to write the heat diffusion equation as
\begin{equation}
T^{n+1}_i \approx T^n_i+\frac{\alpha \delta t }{\delta r}\left[\frac{T^{n}_{i+1}-T^n_i}{r_i}+\frac{T^{n}_{i+1}-2T^n_i+T^n_{i-1}}{\delta r}\right ] \label{thermabl}
\end{equation}
where subscripts and superscripts mark spatial steps and time steps, respectively.

Thermal ablation depends not only on conductive heat transfer within the bolide $\Omega$ (Equation (\ref{one-d heat equation})) but also on the efficiency of heat transfer from the surroundings to the bolide surface. We model this by a relation analogous to Newton's law of heating, matching the heat gain through the bolide boundary with the conductive heat flux into the bolide interior. Thus we impose a Robin boundary condition (a weighted sum of Neumann and Dirichlet boundary conditions):
\begin{equation}
\kappa \frac{\partial T}{\partial r} =\aleph (T_S-T),
\end{equation}
where $\kappa$ is the thermal conductivity of the impactor and $T_S$ is the ambient temperature of surrounding fluid. The proportionality constant $\aleph$ is a property of both the surrounding fluid and the planetesimal and can be described as $\aleph = \kappa_{S} A_b/V_b= 3\kappa_S/r_b$, where $r_b$ is the total radius of the bolide at an instantaneous time. This is implemented by requiring that
the temperature at the bolide outer edge (i.e. at radial grid point
$n_{\mr{out}}$) is given by:
\begin{equation}
T_{n_{\mr{out}}} \approx \frac{(\aleph'T_S+T_{n_{\mr{out}}-1})}{1+\aleph'} \label{robinBC}
\end{equation}
where $\aleph' \equiv \aleph \delta r/ \kappa$ and $\delta r$ is the radial grid spacing at the bolide
outer edge. \citet{french12} compute thermal conductivities $\kappa_S$ for a H-He mixture for a Jovian-analogue. As is discussed in their work, calculations of transport properties in general and thermal conductivities in particular involve DFT-MD with applications of linear response theory \citep{kubo57} and is more complex than mere DFT-MD EOS calculations. The most general thermal conductivity is a combination of electronic and ionic parts, $\kappa_e$ and $\kappa_i$, and shows marked differences in contribution in Jovian envelopes. Where electrons are bound and thermal energy is transported by kinetic energy loss through intermolecular collisions, $\kappa_e$ is expected to be sub-dominant. The conductive transport of heat dominates quite deep in Jovian structures where metalised hydrogen becomes abundant. We use the \citet{french12} combined contributions to the thermal conductivity for the Jupiter-averaged helium fraction of $\overline{Y} \approx 0.275$. 
    
For a unique solution, our Dirichlet boundary conditions are $T(r=0,t)=0$, $T(r=R_{pl},t=0)=170\, \mr{K}$, with $T(r=R_{pl},t)$ given by the Robin condition of Equation (\ref{robinBC}). Where and when the temperature of an evolved region exceeds the melting temperature, we enforce a reduction in the mass and radius such that the material ablates accordingly. There are three timescales with which care must be applied in consideration of our problem.  The first, $dt$, is the timescale for differencing the coupled equations describing the deceleration and frictional ablation of the bolide (Equations (\ref{y1}-\ref{y6})).
The second, $\delta t$, is the timescale for differencing the thermal diffusion equation of the bolide (Equation (\ref{thermabl})), whilst the third, $t_{\mr{diff}} = \delta r^2/\alpha$, is the characteristic timescale on which heat diffuses across a single radial grid cell of the bolide $dr$. Numerical stability in the integration of Equation (\ref{thermabl}) implies the condition $\delta t = k t_{\mr{diff}}$ where $k<0.5$;  Equations (\ref{y1}-\ref{y6}) may  be differenced on a longer timescale and in general we have the ordering: $\delta t < 0.5 t_{\mr{diff}} < 0.5 dt$ (i.e. we have multiple timesteps describing the thermal evolution of the bolide between updates of the bolide position, velocity, etc.).

One may want an adaptive time (i.e., varying $dt$) schema to minimize the expensive computation times taken for the boundary temperature to become sufficient for thermal ablation to begin operating. We used a simple straightforward adaptive scheme for which $dt=0.3\, \mr{s}$ but is increased to 3 seconds when $175\,\mr{K} \le T_{n_{\mr{out}}} \le 1600\,\mr{K}$ to expedite computational time. We have chosen these bounds for a couple of reasons. First, they are temperatures situated intermediately between the effects of frictional ablation (which initially reduces the temperature of the bolide surface by atomising material) and turn-on of thermal ablation (which occurs at 1811 K). Second, they ensure that $dt$ is sufficiently small that the initial frictional peaks in the upper outer envelope are finely evaluated with the condition that $dt>t_\mr{diff}$. We have checked our adaptive scheme with the non-adaptive (i.e., constant $dt$) schema and these yield precise comparisons for equal input parameters and diffusion timescales. The condition $dt>t_\mr{diff}$ implies that the resolution within a planetesimal must vary with the planetesimal size $R_{pl}$, with the number of grid cells defining a resolution size $dr$ determined from the condition $n_{r_2} \approx n_{r_1}\,(R_{pl_2}/R_{pl_1})$. For a 30 metre bolide, we use an $n_r$ value of 15,000, such that a 1 km bolide will have $n_r=500,000$ grid numbers. Input for the normalizing value $k$ is also necessary. A natural question is whether choices of $n_r$ and $k$ yield equivalent results. We ran models for a host of bolide radii from 30 metres to 1 km for widely separated interior resolutions $n_r$ relevant for each and found very good agreement between all values of $k$ for fixed $n_r$ and all $n_r$ values for any individual $k$. We choose a $k$ of 0.4 for fastest computations for all runs throughout our work.

\section{Results}\label{results}

We explore three main themes in our results. First, we discuss a representative case which serves to show the bimodal nature of the ablation process. Second, we probe the variation of impactor angle, impact velocity, and initial impactor radius for four different compositions to explore the rich diversity of scenarios and establish iron as the most conservative case amongst all our bolide compositions. We then narrow our attention to parameter variations for pure iron bolides.

\begin{figure*}
\subfloat{\includegraphics[width = 0.6\textwidth]{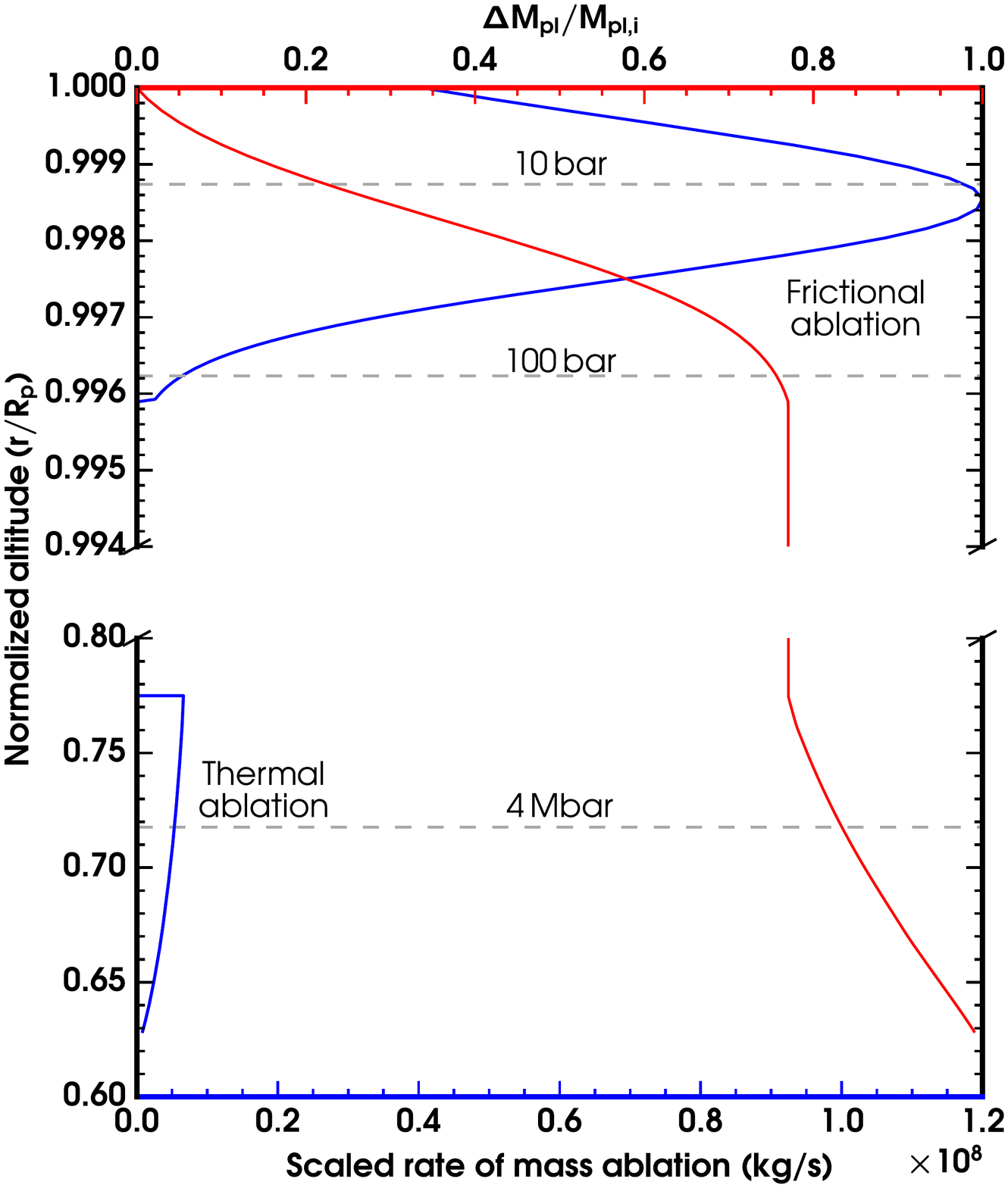}}\\
\vspace{-1.5cm}
\subfloat{\includegraphics[width = 2.1in, height=2.1in]{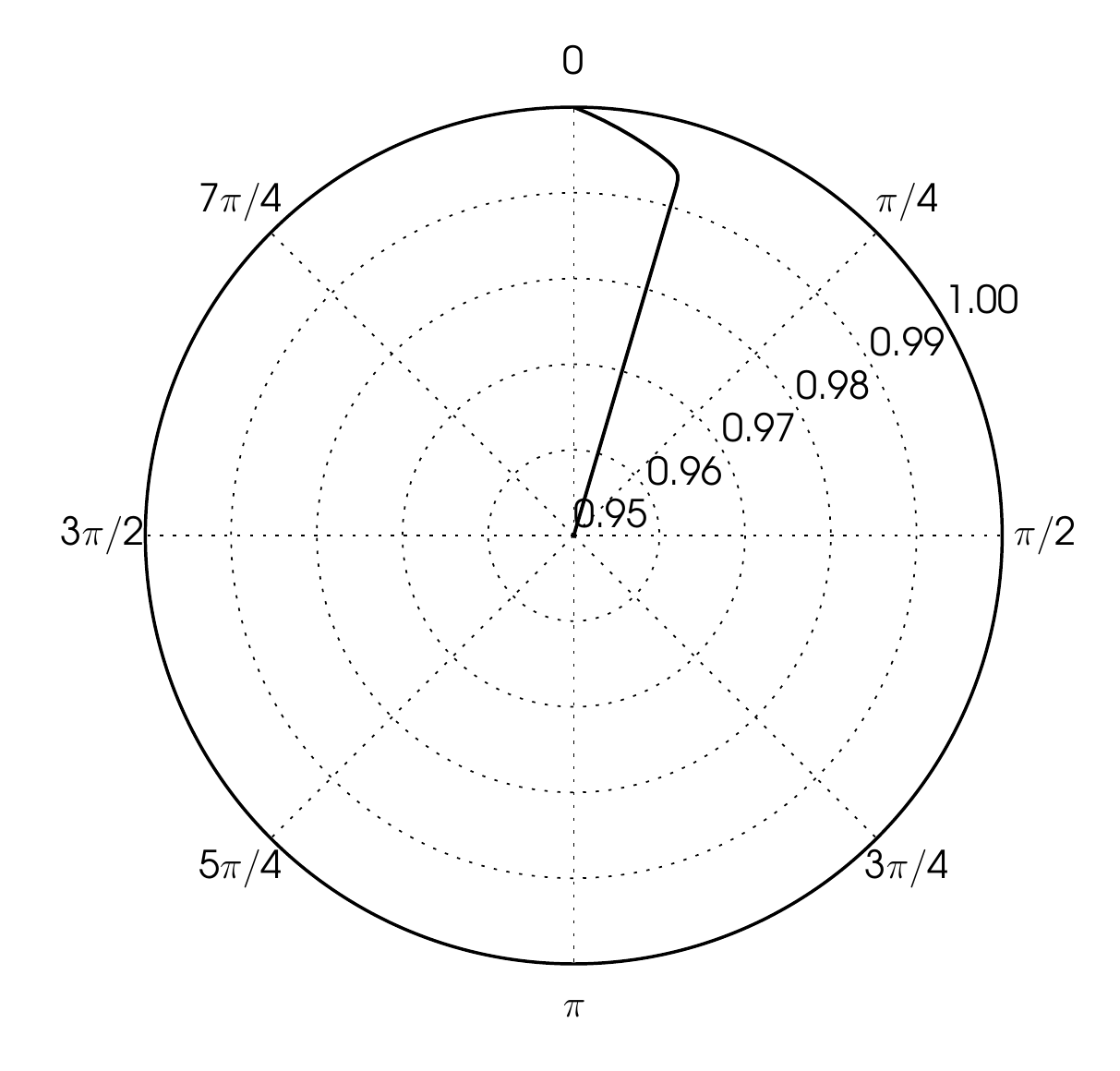}} 
\subfloat{\includegraphics[width = 2.1in, height=2.in]{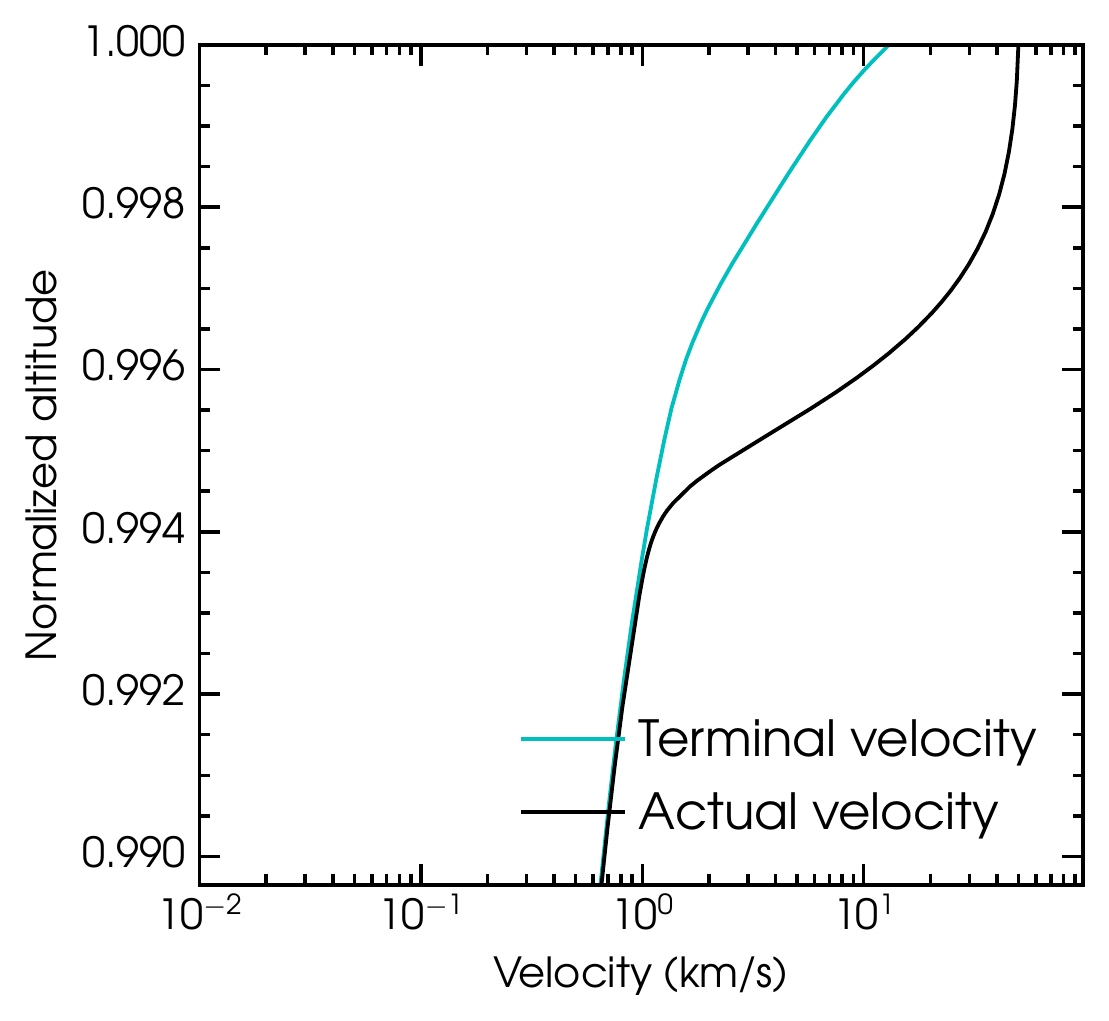}}
\subfloat{\includegraphics[width = 2.1in, height=2.5in]{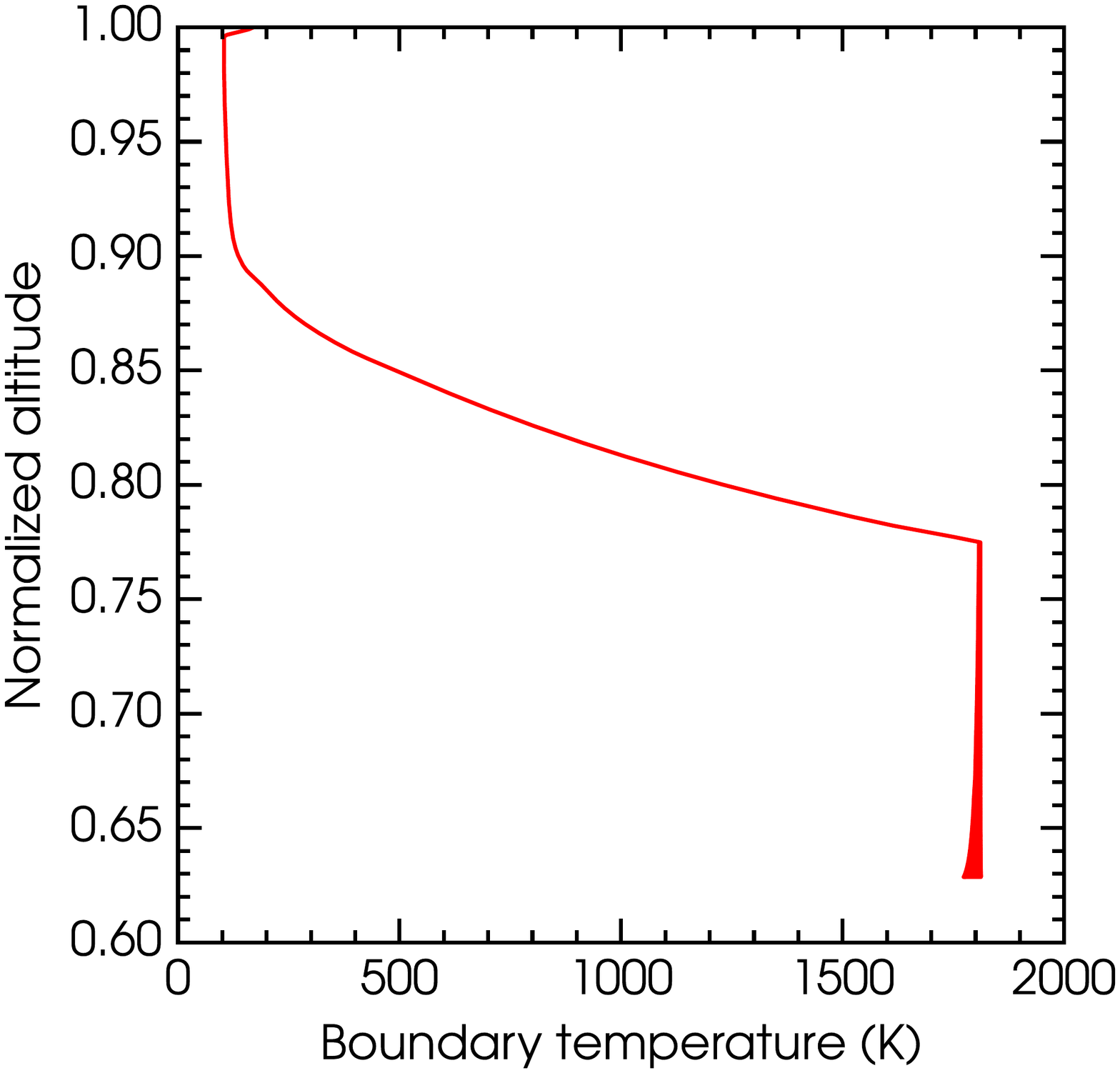}}
\caption{Planetesimal mass ablation and dynamics for an iron bolide with initial radius of 30 metres, impact velocity of $50 \,\mr{km\,s^{-1}}$, and entry angle of 45 degrees. Top main panel: the scaled rate of mass ablation (bottom axis, shown by blue curve) and corresponding fractional mass ablation (top axis, shown by red curve) of the impacting planetesimal as a function of normalised altitude  showing the characteristic bimodal feature of planetesimal ablation. The top and bottom blue distributions respectively show the frictional and thermal rates of mass ablation. Horizontal gray dashed curves denote certain isobars. For ease of visualisation, the rate of thermal ablation has been scaled by a factor 30. Bottom left panel: the dynamical trajectory of the planetesimal in polar coordinates. The azimuthal drag force always acts towards driving the bolide along a purely radial path. The angular values are inflated by a factor of 50 for visibility, whilst the true maximal angle is 0.34 degrees. Bottom middle panel: Actual velocity (black curve) and terminal velocity (cyan curve) evolution as a function of normalized altitude, showing convergence of the two velocities by the time the bolide has traversed  1\% of the planet's radius, which corresponds to the point where the bolide trajectory becomes radial, $\dot{\theta} \rightarrow 0$ (bottom left panel). Bottom right panel: the boundary temperature of the planetesimal as a function of planetary altitude.}
\label{fig:bimodalityillustration}
\end{figure*}

\subsection{Case Study}
We show a typical dynamical evolution featuring the bimodal ablation process in Figure~\ref{fig:bimodalityillustration} for an iron planetesimal of impact radius $R_{pl}=30$ m, initial velocity $|\dot{\vec{r}}|=50\,\mr{km\,s^{-1}}$, and angle of impact $\phi_i=45$ degrees. All of our results in the following sections are run until the impactors have 1 percent of their total mass left, by which point we consider them to be completely ablated, or until contact with the core-inner envelope boundary at  $\sim 41$ Mbar. The top main panel in Figure \ref{fig:bimodalityillustration} shows that frictional ablation initiates upon impact at 1 bar with a mass ablation peak at 10 bar of $2 \times 10^{8}\mr{\,kg\,s^{-1}}$, persisting just beyond 100 bar. This ablation process is generally efficient, as may be seen from the red curve showing the fractional mass ablation along the path, $1-M_{pl_\mr{f}}/M_{pl_\mr{i}}$; nearly 80 percent of the mass of the bolide has vaporised upon completion of frictional ablation. What remains of the impactor then gradually heats up (through Equation (\ref{robinBC})), rising from the boundary temperature left by the last instance of frictional ablation; the rise in temperature is however slow in the upper outer envelope because $\aleph$' is exceptionally small due to the small thermal conductivities, $\kappa_S$, in the envelope. The planetary envelope attains a temperature in excess of the melting temperature of the bolide at 3176 bar, but the
bolide surface does not warm to this temperature until the
bolide has penetrated to beyond 1 Mbar. Thermal ablation completes the planetesimal's wholesale ablation.

The evolution of the bolide surface temperature is shown in the bottom right panel of Figure \ref{fig:bimodalityillustration} and shows that this is close to the melting temperature at normalised altitudes of $<0.8$ (note that the small finite temperature spread in this region is
 an artifact of finite grid resolution since as each layer of the bolide is evaporated, a finite time is taken for the slightly colder material in the adjacent cell to be heated to the melting temperature).

 The polar dynamical path of the bolide is shown in the bottom left panel of Figure~\ref{fig:bimodalityillustration}. The planetesimal enters the outer envelope at 45 degrees to the NJP and the persistent azimuthal aerodynamic force of Equation (\ref{y5}) acts to divert its early path. More particularly, the deceleration of $\theta$ in Equation (\ref{y5}) becomes increasingly less negative as $\dot{\theta}$ decreases thereby directing $\theta$ to invariance of a purely radial trajectory. The angular values in the figure are inflated by a factor of 50 for visibility whereas the true maximal angle is about 0.006 radians or 0.34 degrees.

 The bottom middle panel shows the variation of the actual velocity (black curve) against the terminal velocity (cyan curve) as a function of normalised altitude. The terminal velocity monotonically decreases along the bolide's path due to an increasing planetary atmosphere density. The two velocities converge at a normalised altitude of $\sim 0.99$ which is also the point at which the trajectory becomes more or less radial (bottom left panel). The bolide's continued descent in the lower middle panel of Figure \ref{fig:bimodalityillustration} is at constant radius; the bolide remains at terminal velocity which decreases mildly along the trajectory due to increasing atmospheric density. Thermal ablation then completes the bolide's destruction.

\begin{table*}
\begin{center}
\begin{tabular}{||c c c c||}  
\hline
Planetesimal Composition & Density ($\mr{kg\,m^{-3}}$) & $Q_{abl}$($\mr{MJ\,kg^{-1}}$) & Yield Strength ($\mr{MN\,m^{-2}}$)\\
\hline
Iron & 7800 & $8.26$ & 100\\
Rock & 3400 & $8.08$ & 10\\
Carbonaceous & 2200 & 5.00 & 1\\
Ice & 1000 & $2.80$ & 1\\[.5ex]
\hline
\end{tabular}
\caption{Planetesimal material properties: density, specific sublimation heats, and yield strengths for iron, rock, carbonaceous, and ice planetesimals used throughout this work. The latter two are critical for deposition locations of ablated material. The data are sourced from \citet{chyba93}, \citet{podolak88}, and \citet{petrovic03}.}
\label{properties}
\end{center}
\end{table*}

\subsection{Initial Condition Variations with Frictional Ablation for different Planetesimal Compositions}\label{ic variations}

With our fiducial model we may vary a range of parameters such as impactor composition, entry radius $R_{pl}$, initial velocity $|\dot{\vec{r}}|$, and incident angle $\phi_i$. The output is an illustrative compendium showing planetesimals' fractional mass ablation within the outer envelope. The putative bolide components of the SL9 event used in \citet{korycansky06} were of 1 km in diameter, impact velocities of 61 $\mr{km\,s^{-1}}$, and $\phi_i=43.09$ degrees, where the last two are computed from the arithmetic means of the 21 cometary fragments. Our following sweeps use a pseudo-representative case of 1 km in radius, impact velocity of 30 $\mr{km\,s^{-1}}$, and $\phi_i=45$ degrees. The chosen radius value is in keeping with the idea of `worst-case', where for the velocity we use half the SL9 mean to also be representative of the lowest value of our fiducial set of 10 $\mr{km\,s^{-1}}$. The chosen impact angle is representative of the averaged SL9 components. Whatever two parameters in the following investigations are not swept over are fixed at the two corresponding values from above. Material properties used for the model runs over ICs are shown in Table \ref{properties}.

\subsubsection{Planetesimal Initial Velocity}\label{planetesimalinitialvelocity}

\begin{figure*}
\begin{center}
\includegraphics[width=\textwidth]{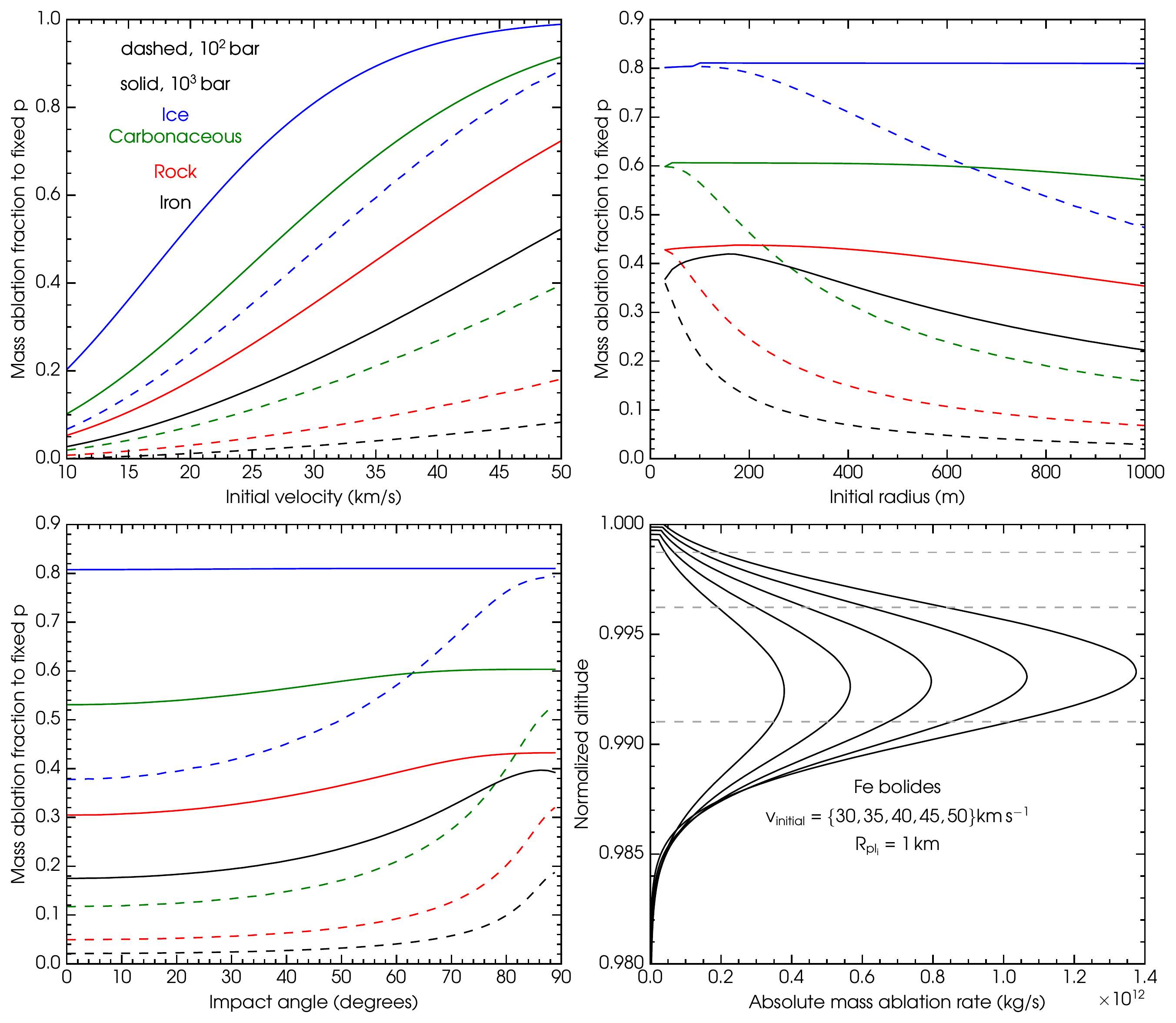}
\end{center}
\caption{Variations in initial parameters for frictional ablation alone. Top left panel: fraction of mass ablation to 100 bar and 1000 bar for variations in the initial velocity of a 1 km impactor incident at 45 degrees to the NJP. As indicated by the legend, dashed curves represent an isobar of 100 bar and solid curves 
represent an isobar of 10$^3$ bar. Blue, green, red, and black curves respectively symbolise ice, carbonaceous, rock, and iron planetesimals. Top right panel: mass ablation fraction of material vaporised to the 100 bar  and 10$^3$ bar levels as a function of initial planetesimal scale for a $30\, \mr{km\,s^{-1}}$ bolide incident at 45 degrees to a zenith. Colors and line styles are equivalent to those in the top left panel. Bottom left panel: the mass ablation fraction of material vaporised to 100 bar and 1000 bar as a function of planetesimal entry angle for a $30\, \mr{km\,s^{-1}}$ bolide with initial radius of 1 km. The color and line style patterns are equivalent to the top panels. Bottom right panel: the rate of mass ablation for iron planetesimals with entry angles of 45 degrees, impact radii of 1 km, and $v_\mr{{initial}}=\{30, 35, 40, 45, 50\}\,\mr{km\,s^{-1}}$, left to right. Light gray dashed lines indicate isobars of 10, 100, and 1,000 bar.}
\label{fractional mass ablations}
\end{figure*}

Velocity variations are important for a planetesimal's evolutionary fate due to the cubic power of initial velocity in the mass ablation in Equation (\ref{mpldot}). Considered velocities are $v_\mr{{initial}} \in [10,50]\, \mr{km\,s^{-1}}$ in light of the impactor values from the SL9 \citep{harrington04} and Tunguska \citep{chyba93} events.\footnote{The average radial velocity of SL9 components was about $44\,\mr{km\,s^{-1}}$ \citep{harrington04} whilst the median of near-Earth asteroids is near $15\, \mr{km\,s^{-1}}$ \citep{chyba93}.} We recognise that impactor velocities less than the escape velocity of Jupiter are problematic unless one proposes a mechanism by which the objects might already have become bound to the planet. With this acknowledgement, we consider how velocity changes modify the fractional mass ablation of planetesimals, 1-$M_{pl_f}/M_{pl_i}$, where $M_{pl_i}$ is the mass at impact and $M_{pl_f}$ is the planetesimal mass at some final isobar. The top left panel of Figure~\ref{fractional mass ablations} shows the fractional mass ablations for a bolide of 1 km incident at $\pi/4$ radians in the outer envelope to the 100 and 1,000 bar levels as a function of initial impactor velocity for different compositions. We see the same qualitative tendency in all our curves or compositions: greater velocities are accompanied by higher fractional mass ablations. Kilometer-size highest velocity (i.e., $\gtrsim 45\,\mr{km\,s^{-1}}$) solid $\mr{H_2O}$ impactors evaporate completely by $10^3$ bar in the upper outer envelope due to the complementary effects of two phenomena: the extended aerodynamic drag dominance (due to ice's low yield strength) that effectively turns on Equation (\ref{y3}) for longer and ice's comparatively low specific sublimation energy. Carbonaceous, rock, and iron bolides ablate sequentially less due to more robust material bonding and higher values of $Q_{abl}$. It has been emphasized that organic impactors will perish when entering the Earth's atmosphere at more than $10\, \mr{kms^{-1}}$ \citep{chyba90}. For our Jovian planet, 1 km carbonaceous impactors with initial velocities greater than $30 \,\mr{km\,s^{-1}}$ lose more than half their mass by $10^3$ bar. Finally, whereas highest velocity ice planetesimals fully ablate before $10^3$ bar, iron bolides of similar velocities lose half their initial mass at this level.

\subsubsection{Impactor Radius}

We may also probe the impact of the initial bolide size. The top right panel of Figure~\ref{fractional mass ablations} shows the effect of initial planetesimal radius on the mass ablation fraction as planetesimals flow through $10^2$ bar and $10^3$ bar. The distribution is flat for small radii but declines at large radii, particularly at low pressures. This is elegantly derived by noting that once the bolide describes a radial trajectory at terminal velocity (see lower middle panel of Figure 3) its velocity is
given by 

\begin{equation}
    |\dot{\vec{r}}_t|^2=\dot{r}_t^2=\frac{2gM_{pl}}{C_D \rho S} \label{term vel2}.
\end{equation}
Substituting this velocity into Equation (8)
and linking the time and spatial derivatives of the differential mass loss rate by the terminal velocity we then obtain: \begin{equation}
\frac{dM_{pl}}{d\vec{r}_t}\dot{\vec{r}}_t \propto |\dot{\vec{r}}_t|^3 \frac{S}{Q_{abl}} \end{equation}  
and thus we can write this equation as

\begin{equation}
    \frac{dM_{pl}}{M_{pl}} \propto \frac{g(r)dr}{Q_{abl}} \label{fractional ablation at terminal}.
\end{equation}
 
The equation illustrates that for a planetesimal of a given composition that asymptotes the true terminal Stokes speed, the mass ablation fraction at some altitude is a constant. Bolide compositions with small characteristic heats of ablation $Q_{abl}$ undergo greatest mass ablations relative to those with higher $Q_{abl}$. The flat relations shift up and down according to the value of $Q_{abl}$. Indeed, analysis of the top right panel shows that icy bolides of all radii studied attain terminal velocity by $10^3$ bar whereas larger rocky and iron bolides do not. The contrast between the size-dependent ablation for small bolides at $10^2$ bar and the generally flat behaviour otherwise is a manifestation of the effect of bolides attaining terminal velocity in the latter cases.

 \subsubsection{Income Angle Variations}\label{income angle variations}
 
The last parameter variation is with initial impact angle $\phi_i$. Note that the requirement that the bolide has some aerodynamical interaction with the planet implies that we can only treat $\phi_i$ values that approach $\pi/2$ asymptotically; given our planetary grid resolution we can however treat the case that $\phi_i$ approaches $\pi/2$ to within $0.7$ degrees. Therefore all references to `90 angle' incidences in this work imply computations carried out at 89.3 degrees. Equation (\ref{fractional ablation at terminal}) is generally valid for terminal-velocity planetesimals and likewise applies to analysis of the bottom left panel of Figure \ref{fractional mass ablations}. We see that the mass ablation fraction is only weakly dependent on impact angle apart from the case of nearly tangential entry. By $10^3$ bar, 1 km bolides of ice sublimate 80 percent of their mass for all impact angles whilst carbonaceous chondrites retain about 45 per cent of the initial mass. Whereas ice and carbonaceous species of the highest angular momenta reach terminal velocities by 500 bar, rock and iron bolides' retention of much of their mass delays their attainment of terminal velocity and this is
demonstrated by a general lack of horizontal features in their curves.  

\subsubsection{Mass Ablation Rate}

The bottom right panel of Figure~\ref{fractional mass ablations} shows the rates of mass ablation for iron bolides of different initial velocities for $\phi_i=\pi/4$ radians and $R_{pl_i}=1\,$km. We notice from the figure that iron planetesimals of markedly different velocities reach peak frictional vaporisation rates at practically the same altitude in the envelope. This observation turns out to be a specific case of a generalisation, such that for some given material composition, the rate of mass ablation peaks at the same $r$ for markedly varying initial velocities. The results of Figure~\ref{fractional mass ablations} illuminate two important generalities amongst the details. First, frictional ablation is efficacious in ablating planetesimals below $10^3$ bar especially for the smallest and weakest of materials. We shall see in Section~\ref{iron explorations} that thermal ablation acts more persistently. Second, iron planetesimals are shown to be `worst-case' scenarios. The figures show that for all IC variations in initial impact velocity, initial impact radius, and angle of entry, iron planetesimals show least normalized mass losses to any planetary altitude compared with ice planetesimals. Moreover, as already noted (solid black curve, upper right panel of Figure 4) we find that
the fractional mass ablation is higher for smaller bodies since iron bolides have not attained terminal velocity by 1000 bar and therefore the dependence of ablation rate on $R_{pl}$ is mainly driven by the $R_{pl}$ dependence of the cross section rather than the velocity in Equation (8). 

\subsection{Exploration of Iron Planetesimals}\label{iron explorations}
\begin{figure*}
\begin{center}
\includegraphics[scale=0.83]{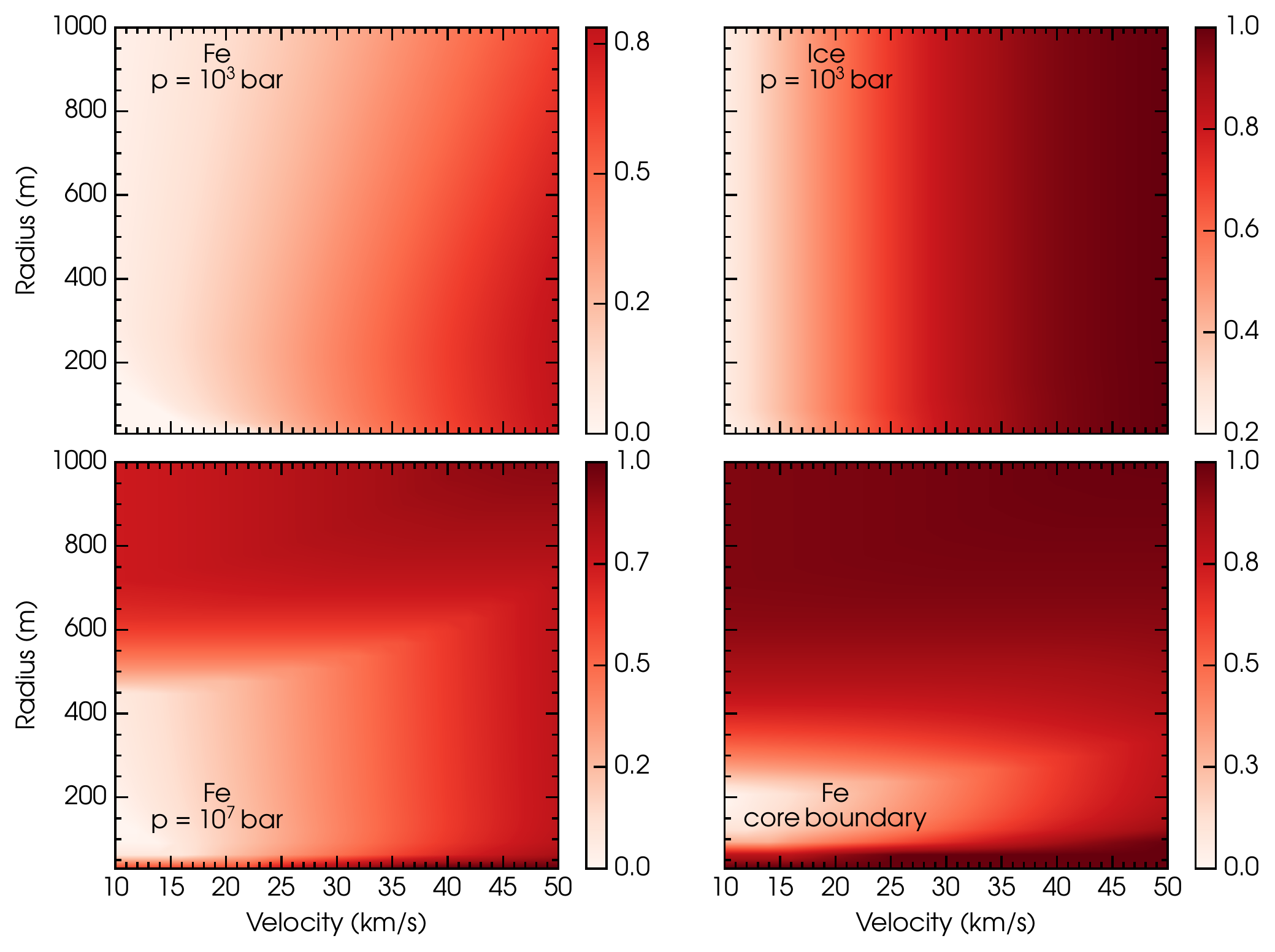}
\caption{Mass ablation fractions of iron and ice planetesimals to different isobars in the Jovian structure for an entry angle of 60 degrees. Upper left panel: iron planetesimals' fractional mass ablation to a pressure level of $10^3$ bar as a function of initial radius and initial velocity. Upper right panel: the same for ice considering only frictional ablation. Lower left panel: iron planetesimals' fractional mass ablation to a pressure level of 10 Mbar as a function of initial radius and initial velocity. Lower right panel: the same as the lower left panel but for the core-inner envelope boundary.}
\label{fig:iron_phi60_contours}
\end{center} 
\end{figure*}

We proceed with the computation of iron impactor evolutions due to frictional and thermal ablation, in which the depth of penetration of planetesimals into the envelopes depends mainly upon their size and velocity. Figure~\ref{fig:iron_phi60_contours} shows two-dimensional mass ablation fraction states and evolutions for pure iron planetesimals to $10^3$ bar, $10^7$ bar, and $4.1687 \times 10^7$ bar in the Jovian envelope structure along with a comparison to icy planetesimals at $10^3$ bar. Each figure shows the fractional mass ablation integrated from the 1 bar level to the different pressure levels. We note that all impact angles for iron show qualitatively similar patterns with one another at every pressure; for clarity we thus restrict ourselves to a discussion of the $\phi_i=60$ degrees case. The arguments to be made about this case will generalize to all representative angles; we show contour plots for $\phi_i=\{0, 30, 60, 90\}$ degrees in Appendix \ref{appendixa} for a complete reference.

Frictional ablation is prominent in upper strata of the outer envelope for almost all planetesimals considered whilst thermal ablation is important well inside the planetary structure in regions where the ambient temperatures of the envelope are greater than the melting point of the planetesimal material and is active only for the smallest of planetesimals. For our planetary structure of \citet{becker14}, the earliest thermal ablation can turn on for iron at 1811 K is at a normalized altitude of 0.986 corresponding to 3176 bar; the low $\kappa_S$ of the outer envelope, however, initiate thermal ablation only towards the bottom of the outer envelope and below. 

Therefore only frictional ablation contributes to the upper left panel of Figure~\ref{fig:iron_phi60_contours}. The $10^3$ bar contour for iron shows mass ablation fractions increasing with increasing initial velocity and decreasing radius. This behaviour follows from the fractional mass ablation equation, $\dot{M}_{pl}/M_{pl} \propto |\dot{\vec{r}}|^3 R_{pl}^{-1}$. Planetesimals of small radii actually have higher fractional rates of ablation than bigger planetesimals for given velocity, with higher velocities showing greater mass ablation fractions for fixed planetesimal size.

Generally, we find that only the highest velocity iron bolides having smallest initial radii efficiently ablate about 60 to 80 percent of their mass to the $10^3$ bar level. In progressing from 1 bar to $10^3$ bar, the mass ablation fraction increases for bigger radii and smaller velocities. The $10^3$ isobar level is significant because it is here that constancy in $\theta$ occurs for the great majority of planetesimals such that $\dot{\theta}=0$, forcing planetesimals into purely radial kinematics. The fact that about 80 percent of the mass is ablated by $10^3$ bar is of great significance; the fully convective nature of the envelopes implies that deposited material can mix upwards to the 1 bar surface on short timescales.

The upper right panel of Figure~\ref{fig:iron_phi60_contours} shows the analogue of the upper left panel but for pure ice planetesimals and only incorporates frictional ablation. Similarly to iron, all impact angles for ice show the same behaviour by $10^3$ bar; we therefore consider $\phi_i=60$ degrees for comparison with the upper left panel. We find that 80 percent or more of the icy impactor mass is removed for impact velocities of 30 $\mr{km\,s^{-1}}$ or greater. As discussed in Section \ref{planetesimalinitialvelocity}, there are two parameters that explain the resulting fragility of icy objects: $Q_{abl}$ and the yield strength. A lower $Q_{abl}$ for ice allows for higher mass ablation rates for given radius and velocity, whilst a lower yield strength allows frictional ablation to persist longer than for an iron bolide with initially equivalent radius, velocity, and impact angle.

The lower left panel of Figure~\ref{fig:iron_phi60_contours} shows the 10 Mbar level contour in the upper inner envelope which lies well beyond the level where thermal ablation can in principal affect planetesimals. When compared with the upper left panel we see that low-velocity, large-radius planetesimals have ablated significantly between the $10^3$ bar and 10 Mbar level. Bolides with initially lowest velocities and radii remain comparatively unaffected in between these pressure levels.  The inefficient ablation for these objects can be understood in terms of Equation (\ref{y1}) and Equation (\ref{y5}) since both have terms which evolve as $R_{pl}^{-1}$ and because the rate of increase of the gravitational acceleration, and therefore the terminal velocity, gradually increases. Large planetesimals therefore have a larger velocity compared to smaller bolides at altitudes where the frictional ablation condition of $F_D/(2S)>\mr{yield\, strength}$ turns on again. We find that low velocity bolides of 500 metres and greater show this character whereby approximately 70 percent or more of the mass is lost due to the resurgence of frictional ablation. Smaller bolides of characteristically lower inertias attain terminal velocities more readily and this reduces the possibility of frictional ablation,
allowing thermal ablation to complete their destruction.

The lower right panel of Figure~\ref{fig:iron_phi60_contours} shows the state of planetesimals at the core-inner envelope boundary. We find that the cavity feature (i.e., the parameter space for which most of the bolide mass is still retained) seen for 10 Mbar is further distinguished from its surrounding parameter space at the core boundary and is confined to a narrower range of initial radii ($\sim$ 90 - 250 m) and rather low initial velocities. Smaller bolides (of $\sim$ 30 m scale) which had not undergone frictional ablation by 10 Mbar are nevertheless thermally ablated between 10 Mbar and the core; larger bolides (of $\sim$ 400 m scale) can eventually
undergo frictional ablation because their terminal velocity increases due to the increasing gravitational field strength at small planetary radii. Only bolides in a very limited size range are large enough that they
don't reach the sublimation temperature and yet small enough for
their low terminal velocities to render them immune to frictional ablation.
Given the uncertainties and limitations (for example, of excluding fragmentation and lateral spread) that enter our modeling, this narrow range may indeed be either somewhat smaller or simply non-existent in a scenario with more modes of ablation.

\begin{figure}
\begin{center}
\includegraphics[scale=0.6]{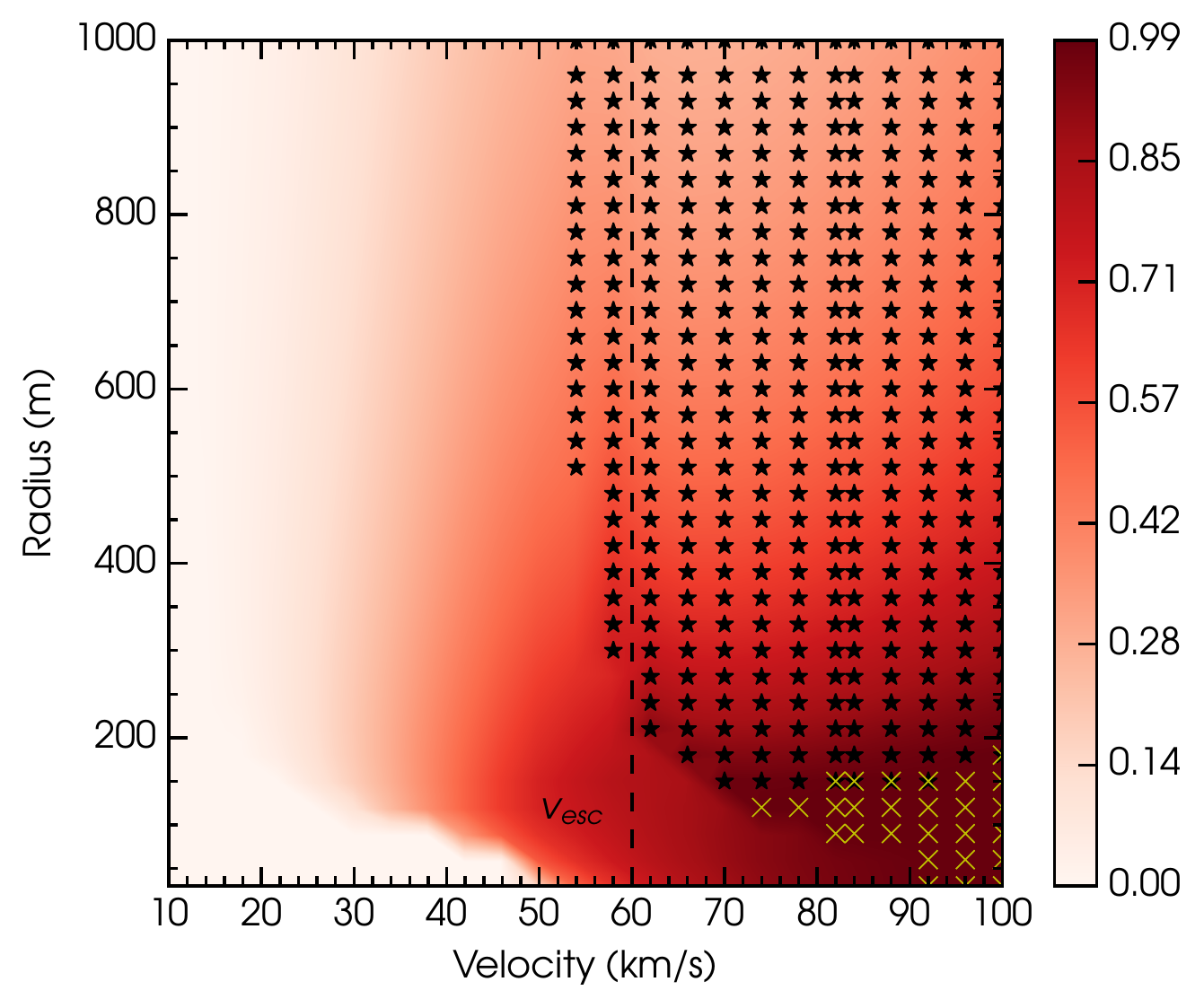}
\end{center}
    \caption{Mass ablation fraction to 126 seconds for right-angle planetesimals with initial velocities exceeding that of Jovian escape. Stars and yellow crosses represent escaped planetesimals and totally ablated planetesimals, respectively. We see that planetesimals of velocities exceeding the Jovian escape should offer chemical depositions that should be observable at the 1-bar level.} 
    \label{fig:highervelocities}
\end{figure}

If impacting planetesimals move with hypervelocities beyond the escape velocity of Jupiter of approximately 60 $\mr{km\,s^{-1}}$, some high $\phi_i$ bolides can ablate and still escape the upper outer envelope beyond a normalised altitude of 1. Figure~\ref{fig:highervelocities} shows the case of bolides impacting at right angles to the local zenith to 126 physical seconds. This time value is ten times the time required for vaporisation of a 30 metre bolide travelling at 100 $\mr{km\,s^{-1}}$ and is a good benchmark time to gauge when all highest-velocity planetesimals either escape the system or completely ablate. Planetesimals capable of escaping the Jovian envelope are represented by black stars, whilst the parameter space which undergoes complete ablation is covered by yellow crosses. Bolides of sizes 30 to 200 metres with $\sim$ 80 $\mr{km\,s^{-1}}$-100 $\mr{km\,s^{-1}}$ completely ablate upon impact, whilst larger bolides with these high kinetic energies manage to escape the Jovian radius (e.g., a 1 km planetesimal retains $\sim 0.60$ of its initial mass in the process). At future times, the region of the plot involving velocities greater than the escape velocity does not evolve whilst the low velocity regime is subject to progressive ablation as in Figure \ref{fig:iron_phi60_contours}. The former region does not evolve because we do not model the re-entry of planetesimals that may fall back to the planet due to their decreased velocities. Planetesimals of velocities exceeding Jovian escape offer chemical depositions that should be observable perhaps even in the atmosphere of such planets given the proximity of the 1 bar level and mixing processes. 

\section{Discussion}\label{discussion}

Our model is a simple description of planetesimal dynamics with implications for the enrichment of Jovian-like exoplanetary envelopes. There is consequently extensive potential for further improving the sophistication of our model and we here discuss a few examples of potential refinements. First, \citet{nellis97} highlights that there is indeed a continual process of molecular dissociation into a metallic plasma; there is thus no real distinct boundary between molecular layers and monoatomic conductive hydrogen and the band-gap separation is bridged continuously with increasing pressure rather than jumping discontinuously. Our utilization of the Becker et al. (2014) Jovian adiabat alleviates this issue, but actuality is clearly more subtle than portrayed here. Hydrogen and helium may not adhere very accurately to the linear-mixing model, such that immiscibility of species resulting in non-linear mixing may be of importance for Jovian planets \citep{becker14,militzer13}. In addition, the model does not assume the formation of helium droplets settling downward from a helium-depleted upper envelope to an inner envelope with higher He abundance due to chemical separation or gravitational settling in a process called demixing \citep{salpeter73}. This effect would release additional energy by added heating through material movement into a deeper potential, acting to place iron planetesimals' ablation higher in the envelopes. 

A more accurate paradigm would include metallic species such as carbon, nitrogen, and oxygen. Another limitation of our paradigm is that a more sophisticated model would require functional thermal conductivities for impactors as a function of temperature, pressure, and density, requiring knowledge of the complete phase diagram of iron conductivities from high-pressure condensed-matter physics. This is currently an extreme challenge. Further, our thermal ablation assumes an invariant melting temperature, thereby providing a host of qualitative features. One would want a set of complete phase diagrams for the varied impactor compositions, detailing the location of freezing points with pressure and temperature variations. In pursuit of this goal, a first model may be to use a Tillotson EOS devised as an analytic thermodynamical model useful for shocks and material ablation of hypervelocity impacts. 

It should prove illuminating to model lateral spread within the framework of frictional vaporisation and thermal melting. In addition to the aerodynamic drag, the impactor as a whole may become compressed and flattened, becoming progressively spheroidal from an initially spherical geometry. This is most prominent in the smallest objects (objects of $\lesssim$ 1-2 km) which often expand significantly horizontally or pancake and for which hydrodynamic instabilities, such as Kelvin-Helmholtz and RT, are minimal \citep{korycansky20}. The impactor becomes distorted once aerodynamic forces overcome the yield strength of the impactor, deforming the bolide laterally due to the difference in the magnitude of the drag force at the impactor's anterior and posterior. The net effect for frictional ablation would be to induce a positive feedback which increases Stokes loading of the impactor due to a growing radius.

Finally, two most important inclusions should be of fragmentation and thermal radiation. These would act to place the ablation at higher levels in the envelopes, perhaps being of such efficiency as to keep all mass loss within the outer envelope. Accurate modelling of the first of these phenomena is complicated and the resulting micro-planetesimals are believed to have non-negligible rotational momenta resulting from catastrophic fractionation, although the latter of these should not be such a problem due to the weakness of Magnus forces on such objects (see our discussion in Section \ref{FA}). It is also still debated whether it may be more realistic to model fractionation of meteoroids through considerations of raw material strength \citep{artemieva01} or by growth rates of hydro-dynamical instability modes \citep{korycansky05}. 

One might wonder about modelling the effects of deposited material from planetesimals upon the planetary structure for subsequently impacting planetesimals. The depositon of more conductive heavy metallic elements would decrease the bulk specific heat capacity in a unit volume originally containing H and He thereby increasing the internal thermal lapse rate. An increased temperature gradient in the planetary structure would induce greater thermal ablation by Equation (\ref{robinBC}) and higher frictional ablation due to larger envelope densities. This therefore reiterates the point that our model is capable of placing lower limits on the efficiency of wholesale ablation of planetesimals compared to models with more detailed physics.

\begin{figure}
\includegraphics[scale=0.55]{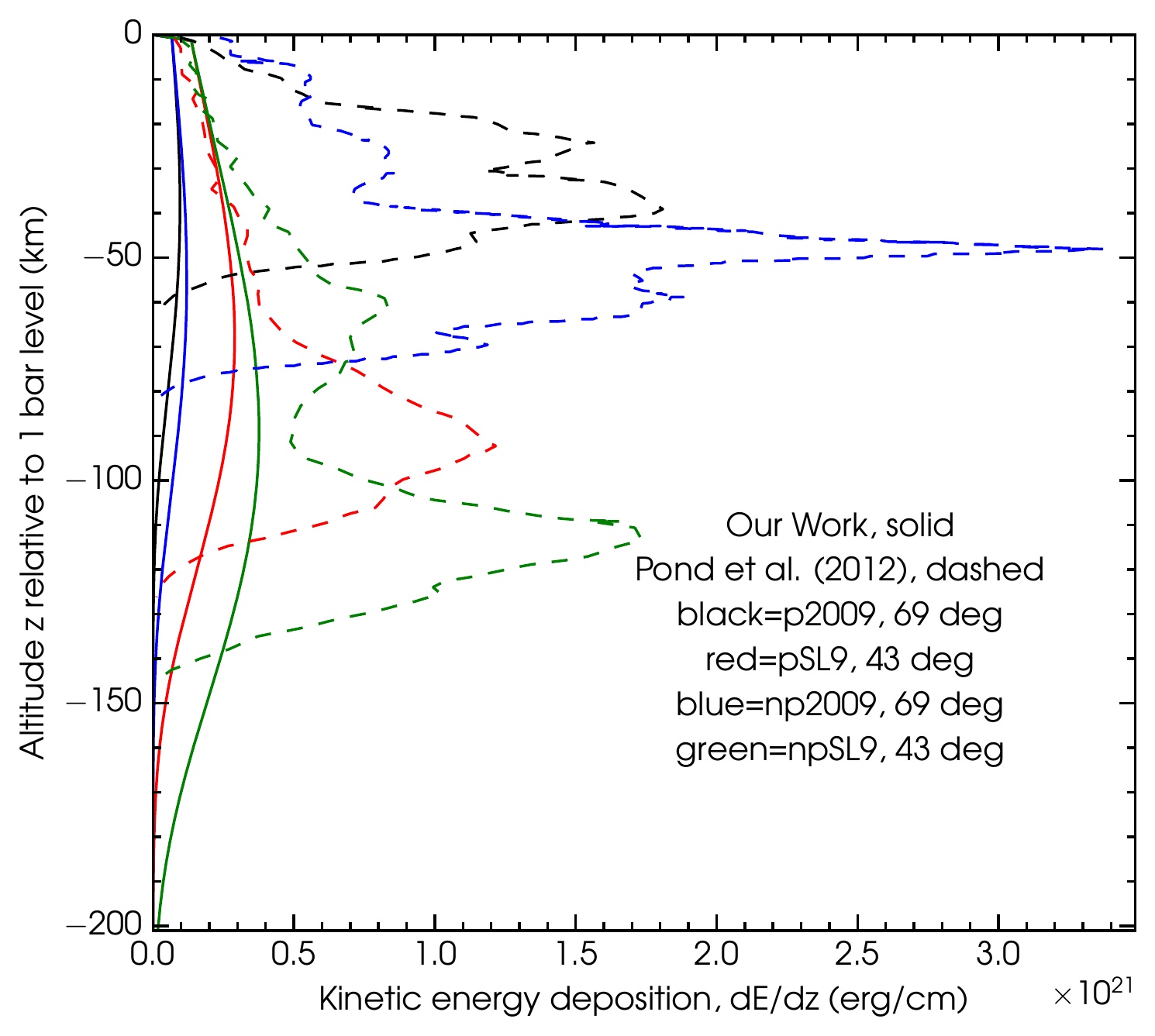}
\caption{Comparison with the work of \citet{pond12} of the $z$-projected altitude relative to the 1 bar level against the kinetic energy depositions.}
\label{fig:Pond_comp}
\end{figure}

We have compared our results with the ZEUS-MP2 hydrodynamical simulations of SL9 and 2009 Jovian impacts of \citet{pond12}; see Figure \ref{fig:Pond_comp}. Our models for these two events with fiducial impact parameters of \citet{pond12} for incident angles, material densities, and planetesimal sizes of putative impactors are reasonably understood when compared with their kinetic energy deposition curves along the canonical $z$ direction (cf. Equation (6) of \citet{korycansky06}). The solid curves show our models whilst the dashed curves are those of Pond et al. (2012). We show four suggestions for these bolides that are most commensurate with the available data, a pair for each case: a porous 2009 impact at 69 degrees to a local vertical (p2009, 69 degrees), a porous SL9 at 43 degrees (pSL9, 43 degrees), a non-porous 2009 impactor at 69 degrees (np2009, 69 degrees), and a non-porous SL9 at 43 degrees (npSL9, 43 degrees). A porous density of 0.6 g/cm$^3$, non-porous density of 0.917 g/cm$^3$, and $R_{pl}=500$ m have been used as in \citet{pond12}. Note that because these are icy bolides, our model only considers the effect of frictional ablation. We find that our p2009-69 degree object ablates 67 per cent of its mass
at the altitude at which the \citet{pond12} results find that such an object is completely ablated; correspondingly we find 88 percent ablation of the pSL9-43 degree object, 69 percent ablation of the np2009-69 degree object and 85 percent ablation of the npSL9-43 degree object at the points where the \citet{pond12} analogues were entirely ablated. These results imply that, as expected, ablation is more efficient in three dimensions since simulations such as ZEUS-MP2 include elements such as lateral spread and fragmentation of planetesimals, both of which serve to place terminal ablation higher in the outer envelope.

We find qualitative similarities with the work of \citet{mordasini15} but cannot make cannot make quantitative comparisons (because of the different planetary structure, incorporated physics, planetesimal composition, and evolutionary stage). They find that planetesimals are completely ablated in the envelope for envelope masses $\gtrsim$30 M$_\oplus$ which is consistent with our finding for our Jupiter-like envelope of mass $\sim$300 M$_\oplus$, for a range of planetesimal sizes and impact velocities. We find an analogy with the work of \citet{mordasini15} that shows that only a limited range of sizes of rocky planetesimals are able to reach the core. In our case this feature is due to these planetesimals being too small to be subject to efficient frictional ablation and too large to be effectively thermally ablated (see Figure \ref{fig:iron_phi60_contours}).

Finally, \citet{podolak88} also find substantial ablation of planetesimals accreting onto protoplanetary envelopes. They computed this in the framework of core-accretion using three mechanisms for the ablation of planetesimals: frictional ablation, fragmentation due to mechanical instabilities, and thermal blackbody radiation. A directly meaningful comparison of their work with ours is not possible as their models consider protoplanets with very different masses and radii and hence a markedly different internal structure compared to ours (i.e., a cold Jupiter). Nevertheless, Figure 6 of \citet{podolak88} shows that an iron planetesimal of initial radius 1 km of zero angular momentum experiences wholesale disruption at a protoplaneto-centric distance of 0.7 $R_J$ (about 2.5 times their core radius), and this for a core mass of 16.8 Earth masses and envelope mass of 27 $M_{\oplus}$ \citep{bodenheimer86}. In our models, the envelope is much more massive ($\sim$300 $M_{\oplus}$) and for a Jovian-like steady-state envelope structure leads to substantial planetesimal ablation.

\section{Conclusions}\label{conclusions}

We have developed a simple model for the frictional and thermal ablations of incident impactors into Jovian-like envelopes. We have found that frictional ablation is dominant in the upper strata of the outer envelope and is generally efficacious. Planetesimal initial velocity, composition, impact angle, and radius are the factors which control the magnitude and location of this mass ablation, with higher (lower) radius, lower (higher) $\phi_i$, and lower (higher) impact velocity leading to comparatively less (more) mass ablation. The inclusion of thermal ablation adds an important phase of ablation deeper in the envelope where the ambient temperature well exceeds the melting temperature of the impactor and when the thermal conductivity is sufficiently large. The effects of thermal ablation become relevant in regions where the velocity asymptotes terminal velocity in the high-pressure regimes beyond about 3100 bar, and in cases of small planetesimals finishes off what frictional ablation began. 

One interesting finding is that for planetesimals of highest velocities and largest radii, aerodynamical loading fully ablates ice bolides by $10^3$ bar and succeeds in ablating around 50 percent of iron bolides' mass. By $10^3$ bar, iron planetesimals of highest velocities having smallest radii are found to ablate about 60 to 80 percent of their mass due to frictional ablation, whilst for ice bolides we find frictional ablation causes planetesimals of all impact angles and radii with initial velocities greater than 30 $\mr{km\,s^{-1}}$ to have lost more than 80 percent of their mass. We find that contact with the inner envelope-core boundary is viable for some planetesimals. A most important finding concerns that of intermediate size, low-velocity planetesimals of $\sim$90 - 250 metres: these  penetrate through our massive envelope and make contact with the core. These middlings are too small to attain the velocities required for frictional ablation and yet too large to be heated to the melting temperature.

One natural motivation of this work is to discover whether the planetesimal depositions may allow telltale detectable features at about the 1 bar level of the atmosphere through convective mixing. Convective circulation timescales in Jupiter are believed to be $t_{\mr{conv}}\sim 3$ years whilst evolutionary times are greater than $10^9$ years \citep{salpeter73}; therefore, the dissolution of planetesimals through the modeled methods should readily remix material on times of order $t_{\mr{conv}}$. Our work therefore has implications for Jovian exoplanetary analogues by using the injected chemistry as an observational tool to help discussions of formation and evolution.  

Our model describes planetesimal ablation and enrichment in the envelope of a Jovian analogue. Our conclusions are therefore most directly applicable to the latter stages of gas-giant formation and evolution. A more comprehensive understanding means that chemical enrichment models through planetesimal ablation must be pursued during the nascent protoplanetary stages when the planets are just forming and are hotter. It is expected that for an initially younger, hotter planet our conclusion that atmospheric chemistry abundance should indeed reflect the time-averaged formation and migration conditions should prove correct. However, proto-gas-giants initially have less massive gaseous envelopes and there may be substantial refractory material which accumulates in the core. In the future, it is therefore important to distinguish beyond which stage in the formation of gas-giants atmospheric chemistry is expected to give a whole description of the chemical enrichment and thus of formation and migration conditions. In addition, future studies can benefit from developing more sophisticated and accurate models to treat the infall of planetesimals which will also include other exoplanetary types. 

\section*{Acknowledgments}
AP extends appreciation to the Gates Cambridge Trust for research funding. This work has been partially supported by the DISCSIM project, grant agreement 341137 funded by the European Research Council under ERC-2013-ADG. We thank Alex Macmillan for initial exploration of this problem in his master's thesis with NM. We thank the referee for their comments and suggestions which have assisted in improving the presentation of our work.




\bibliographystyle{mnras}
\bibliography{ms_0} 

\begin{thebibliography}{}
\makeatletter
\relax
\def\mn@urlcharsother{\let\do\@makeother \do\$\do\&\do\#\do\^\do\_\do\%\do\~}
\def\mn@doi{\begingroup\mn@urlcharsother \@ifnextchar [ {\mn@doi@}
  {\mn@doi@[]}}
\def\mn@doi@[#1]#2{\def\@tempa{#1}\ifx\@tempa\@empty \href
  {http://dx.doi.org/#2} {doi:#2}\else \href {http://dx.doi.org/#2} {#1}\fi
  \endgroup}
\def\mn@eprint#1#2{\mn@eprint@#1:#2::\@nil}
\def\mn@eprint@arXiv#1{\href {http://arxiv.org/abs/#1} {{\tt arXiv:#1}}}
\def\mn@eprint@dblp#1{\href {http://dblp.uni-trier.de/rec/bibtex/#1.xml}
  {dblp:#1}}
\def\mn@eprint@#1:#2:#3:#4\@nil{\def\@tempa {#1}\def\@tempb {#2}\def\@tempc
  {#3}\ifx \@tempc \@empty \let \@tempc \@tempb \let \@tempb \@tempa \fi \ifx
  \@tempb \@empty \def\@tempb {arXiv}\fi \@ifundefined
  {mn@eprint@\@tempb}{\@tempb:\@tempc}{\expandafter \expandafter \csname
  mn@eprint@\@tempb\endcsname \expandafter{\@tempc}}}

\bibitem[\protect\citeauthoryear{{Artemieva} \& {Shuvalov}}{{Artemieva} \&
  {Shuvalov}}{2001}]{artemieva01}
{Artemieva} N.~A.,  {Shuvalov} V.~V.,  2001, \mn@doi [\jgr]
  {10.1029/2000JE001264}, \href
  {http://adsabs.harvard.edu/abs/2001JGR...106.3297A} {106, 3297}

\bibitem[\protect\citeauthoryear{{Atreya} \& {Wong}}{{Atreya} \&
  {Wong}}{2005}]{atreya05}
{Atreya} S.~K.,  {Wong} A.-S.,  2005, \mn@doi [\ssr]
  {10.1007/s11214-005-1951-5}, \href
  {http://adsabs.harvard.edu/abs/2005SSRv..116..121A} {116, 121}

\bibitem[\protect\citeauthoryear{{Becker}, {Lorenzen}, {Fortney}, {Nettelmann},
  {Sch{\"o}ttler}  \& {Redmer}}{{Becker} et~al.}{2014}]{becker14}
{Becker} A.,  {Lorenzen} W.,  {Fortney} J.~J.,  {Nettelmann} N.,
  {Sch{\"o}ttler} M.,   {Redmer} R.,  2014, \mn@doi [\apjs]
  {10.1088/0067-0049/215/2/21}, \href
  {http://adsabs.harvard.edu/abs/2014ApJS..215...21B} {215, 21}

\bibitem[\protect\citeauthoryear{{Benneke}}{{Benneke}}{2015}]{benneke15}
{Benneke} B.,  2015, preprint, \href
  {http://adsabs.harvard.edu/abs/2015arXiv150407655B} {} (\mn@eprint {arXiv}
  {1504.07655})

\bibitem[\protect\citeauthoryear{{Bodenheimer} \& {Pollack}}{{Bodenheimer} \&
  {Pollack}}{1986}]{bodenheimer86}
{Bodenheimer} P.,  {Pollack} J.~B.,  1986, \mn@doi [\icarus]
  {10.1016/0019-1035(86)90122-3}, \href
  {http://adsabs.harvard.edu/abs/1986Icar...67..391B} {67, 391}

\bibitem[\protect\citeauthoryear{{Bronshten}}{{Bronshten}}{1983}]{bronshten83}
{Bronshten} V.~A.,  1983, {Physics of meteoric phenomena}.
D.~Reidel Publishing Company

\bibitem[\protect\citeauthoryear{{Chyba}, {Thomas}, {Brookshaw}  \&
  {Sagan}}{{Chyba} et~al.}{1990}]{chyba90}
{Chyba} C.~F.,  {Thomas} P.~J.,  {Brookshaw} L.,   {Sagan} C.,  1990, \mn@doi
  [Science] {10.1126/science.11538074}, \href
  {http://adsabs.harvard.edu/abs/1990Sci...249..366C} {249, 366}

\bibitem[\protect\citeauthoryear{{Chyba}, {Thomas}  \& {Zahnle}}{{Chyba}
  et~al.}{1993}]{chyba93}
{Chyba} C.~F.,  {Thomas} P.~J.,   {Zahnle} K.~J.,  1993, \mn@doi [\nat]
  {10.1038/361040a0}, \href {http://adsabs.harvard.edu/abs/1993Natur.361...40C}
  {361, 40}

\bibitem[\protect\citeauthoryear{{Crawford}, {Boslough}, {Robinson}  \&
  {Trucano}}{{Crawford} et~al.}{1995}]{crawford95}
{Crawford} D.~A.,  {Boslough} M.~B.,  {Robinson} A.~C.,   {Trucano} T.~G.,
  1995, in Lunar and Planetary Science Conference.

\bibitem[\protect\citeauthoryear{{Field} \& {Ferrara}}{{Field} \&
  {Ferrara}}{1995}]{field95}
{Field} G.~B.,  {Ferrara} A.,  1995, \mn@doi [\apj] {10.1086/175137}, \href
  {http://adsabs.harvard.edu/abs/1995ApJ...438..957F} {438, 957}

\bibitem[\protect\citeauthoryear{{Forbes}}{{Forbes}}{2015}]{forbes15}
{Forbes} J.~C.,  2015, \mn@doi [\mnras] {10.1093/mnras/stv1712}, \href
  {http://adsabs.harvard.edu/abs/2015MNRAS.453.1779F} {453, 1779}

\bibitem[\protect\citeauthoryear{{French}, {Becker}, {Lorenzen}, {Nettelmann},
  {Bethkenhagen}, {Wicht}  \& {Redmer}}{{French} et~al.}{2012}]{french12}
{French} M.,  {Becker} A.,  {Lorenzen} W.,  {Nettelmann} N.,  {Bethkenhagen}
  M.,  {Wicht} J.,   {Redmer} R.,  2012, \mn@doi [\apjs]
  {10.1088/0067-0049/202/1/5}, \href
  {http://adsabs.harvard.edu/abs/2012ApJS..202....5F} {202, 5}

\bibitem[\protect\citeauthoryear{{Harrington}, {de Pater}, {Brecht}, {Deming},
  {Meadows}, {Zahnle}  \& {Nicholson}}{{Harrington}
  et~al.}{2004}]{harrington04}
{Harrington} J.,  {de Pater} I.,  {Brecht} S.~H.,  {Deming} D.,  {Meadows} V.,
  {Zahnle} K.,   {Nicholson} P.~D.,  2004, {Lessons from Shoemaker-Levy 9 about
  Jupiter and planetary impacts}.
Cambridge University Press, pp 159--184

\bibitem[\protect\citeauthoryear{{Helled} \& {Schubert}}{{Helled} \&
  {Schubert}}{2009}]{helled09}
{Helled} R.,  {Schubert} G.,  2009, \mn@doi [\apj]
  {10.1088/0004-637X/697/2/1256}, \href
  {http://adsabs.harvard.edu/abs/2009ApJ...697.1256H} {697, 1256}

\bibitem[\protect\citeauthoryear{{Helling}, {Woitke}, {Rimmer}, {Kamp}, {Thi}
  \& {Meijerink}}{{Helling} et~al.}{2014}]{helling14}
{Helling} C.,  {Woitke} P.,  {Rimmer} P.~B.,  {Kamp} I.,  {Thi} W.-F.,
  {Meijerink} R.,  2014, \mn@doi [Life] {10.3390/life4020142}, \href
  {http://adsabs.harvard.edu/abs/2014Life....4..142H} {4, 142}

\bibitem[\protect\citeauthoryear{{Hueso} et~al.,}{{Hueso}
  et~al.}{2013}]{hueso13}
{Hueso} R.,  et~al., 2013, \mn@doi [\aap] {10.1051/0004-6361/201322216}, \href
  {http://adsabs.harvard.edu/abs/2013A26A...560A..55H} {560, A55}

\bibitem[\protect\citeauthoryear{{Jura}}{{Jura}}{2008}]{jura08}
{Jura} M.,  2008, \mn@doi [\aj] {10.1088/0004-6256/135/5/1785}, \href
  {http://adsabs.harvard.edu/abs/2008AJ....135.1785J} {135, 1785}

\bibitem[\protect\citeauthoryear{{Korycansky}}{{Korycansky}}{2015}]{korycansky15}
{Korycansky} D.~G.,  2015, in Lunar and Planetary Science Conference. p.~1144

\bibitem[\protect\citeauthoryear{{Korycansky} \& {Palotai}}{{Korycansky} \&
  {Palotai}}{2014}]{korycansky14}
{Korycansky} D.~G.,  {Palotai} C.,  2014, in Lunar and Planetary Science
  Conference. p.~1269

\bibitem[\protect\citeauthoryear{{Korycansky} \& {Zahnle}}{{Korycansky} \&
  {Zahnle}}{2005}]{korycansky05}
{Korycansky} D.~G.,  {Zahnle} K.~J.,  2005, \mn@doi [\planss]
  {10.1016/j.pss.2005.03.002}, \href
  {http://adsabs.harvard.edu/abs/2005P26SS...53..695K} {53, 695}

\bibitem[\protect\citeauthoryear{{Korycansky}, {Zahnle}  \& {Law}}{{Korycansky}
  et~al.}{2000}]{korycansky20}
{Korycansky} D.~G.,  {Zahnle} K.~J.,   {Law} M.-M.~M.,  2000, \mn@doi [\icarus]
  {10.1006/icar.2000.6426}, \href
  {http://adsabs.harvard.edu/abs/2000Icar..146..387K} {146, 387}

\bibitem[\protect\citeauthoryear{{Korycansky}, {Harrington}, {Deming}  \&
  {Kulick}}{{Korycansky} et~al.}{2006}]{korycansky06}
{Korycansky} D.~G.,  {Harrington} J.,  {Deming} D.,   {Kulick} M.~E.,  2006,
  \mn@doi [\apj] {10.1086/504702}, \href
  {http://adsabs.harvard.edu/abs/2006ApJ...646..642K} {646, 642}

\bibitem[\protect\citeauthoryear{{Kreidberg} et~al.,}{{Kreidberg}
  et~al.}{2015}]{kreidberg15}
{Kreidberg} L.,  et~al., 2015, \mn@doi [\apj] {10.1088/0004-637X/814/1/66},
  \href {http://adsabs.harvard.edu/abs/2015ApJ...814...66K} {814, 66}

\bibitem[\protect\citeauthoryear{{Kubo}}{{Kubo}}{1957}]{kubo57}
{Kubo} R.,  1957, Journal of the Physical Society of Japan, \href
  {http://adsabs.harvard.edu/abs/1957JPSJ...12..570K} {12, 570}

\bibitem[\protect\citeauthoryear{{Line}, {Zhang}, {Vasisht}, {Natraj}, {Chen}
  \& {Yung}}{{Line} et~al.}{2012}]{line12}
{Line} M.~R.,  {Zhang} X.,  {Vasisht} G.,  {Natraj} V.,  {Chen} P.,   {Yung}
  Y.~L.,  2012, \mn@doi [\apj] {10.1088/0004-637X/749/1/93}, \href
  {http://adsabs.harvard.edu/abs/2012ApJ...749...93L} {749, 93}

\bibitem[\protect\citeauthoryear{{Line}, {Knutson}, {Wolf}  \& {Yung}}{{Line}
  et~al.}{2014}]{line14}
{Line} M.~R.,  {Knutson} H.,  {Wolf} A.~S.,   {Yung} Y.~L.,  2014, \mn@doi
  [\apj] {10.1088/0004-637X/783/2/70}, \href
  {http://adsabs.harvard.edu/abs/2014ApJ...783...70L} {783, 70}

\bibitem[\protect\citeauthoryear{{Mac Low} \& {Zahnle}}{{Mac Low} \&
  {Zahnle}}{1994}]{maclow94}
{Mac Low} M.-M.,  {Zahnle} K.,  1994, \mn@doi [\apjl] {10.1086/187565}, \href
  {http://adsabs.harvard.edu/abs/1994ApJ...434L..33M} {434, L33}

\bibitem[\protect\citeauthoryear{{Madhusudhan}}{{Madhusudhan}}{2012}]{madhu12}
{Madhusudhan} N.,  2012, \mn@doi [\apj] {10.1088/0004-637X/758/1/36}, \href
  {http://adsabs.harvard.edu/abs/2012ApJ...758...36M} {758, 36}

\bibitem[\protect\citeauthoryear{{Madhusudhan}, {Mousis}, {Johnson}  \&
  {Lunine}}{{Madhusudhan} et~al.}{2011}]{madhu11}
{Madhusudhan} N.,  {Mousis} O.,  {Johnson} T.~V.,   {Lunine} J.~I.,  2011,
  \mn@doi [\apj] {10.1088/0004-637X/743/2/191}, \href
  {http://adsabs.harvard.edu/abs/2011ApJ...743..191M} {743, 191}

\bibitem[\protect\citeauthoryear{{Madhusudhan}, {Knutson}, {Fortney}  \&
  {Barman}}{{Madhusudhan} et~al.}{2014a}]{madhu14a}
{Madhusudhan} N.,  {Knutson} H.,  {Fortney} J.~J.,   {Barman} T.,  2014a,
  Protostars and Planets VI, \href
  {http://adsabs.harvard.edu/abs/2014prpl.conf..739M} {pp 739--762}

\bibitem[\protect\citeauthoryear{{Madhusudhan}, {Crouzet}, {McCullough},
  {Deming}  \& {Hedges}}{{Madhusudhan} et~al.}{2014b}]{madhu14c}
{Madhusudhan} N.,  {Crouzet} N.,  {McCullough} P.~R.,  {Deming} D.,   {Hedges}
  C.,  2014b, \mn@doi [\apjl] {10.1088/2041-8205/791/1/L9}, \href
  {http://adsabs.harvard.edu/abs/2014ApJ...791L...9M} {791, L9}

\bibitem[\protect\citeauthoryear{{Madhusudhan}, {Amin}  \&
  {Kennedy}}{{Madhusudhan} et~al.}{2014c}]{madhu14b}
{Madhusudhan} N.,  {Amin} M.~A.,   {Kennedy} G.~M.,  2014c, \mn@doi [\apjl]
  {10.1088/2041-8205/794/1/L12}, \href
  {http://adsabs.harvard.edu/abs/2014ApJ...794L..12M} {794, L12}

\bibitem[\protect\citeauthoryear{{Madhusudhan}, {Ag{\'u}ndez}, {Moses}  \&
  {Hu}}{{Madhusudhan} et~al.}{2016}]{madhu16}
{Madhusudhan} N.,  {Ag{\'u}ndez} M.,  {Moses} J.~I.,   {Hu} Y.,  2016, \mn@doi
  [\ssr] {10.1007/s11214-016-0254-3}, \href
  {http://adsabs.harvard.edu/abs/2016SSRv..tmp...31M} {}

\bibitem[\protect\citeauthoryear{{Marboeuf}, {Thiabaud}, {Alibert}, {Cabral}
  \& {Benz}}{{Marboeuf} et~al.}{2014}]{marboeuf14}
{Marboeuf} U.,  {Thiabaud} A.,  {Alibert} Y.,  {Cabral} N.,   {Benz} W.,  2014,
  \mn@doi [\aap] {10.1051/0004-6361/201322207}, \href
  {http://adsabs.harvard.edu/abs/2014A26A...570A..35M} {570, A35}

\bibitem[\protect\citeauthoryear{{Militzer} \& {Hubbard}}{{Militzer} \&
  {Hubbard}}{2013}]{militzer13}
{Militzer} B.,  {Hubbard} W.~B.,  2013, \mn@doi [\apj]
  {10.1088/0004-637X/774/2/148}, \href
  {http://adsabs.harvard.edu/abs/2013ApJ...774..148M} {774, 148}

\bibitem[\protect\citeauthoryear{{Mordasini}, {Molli{\`e}re}, {Dittkrist},
  {Jin}  \& {Alibert}}{{Mordasini} et~al.}{2015}]{mordasini15}
{Mordasini} C.,  {Molli{\`e}re} P.,  {Dittkrist} K.-M.,  {Jin} S.,   {Alibert}
  Y.,  2015, \mn@doi [International Journal of Astrobiology]
  {10.1017/S1473550414000263}, \href
  {http://adsabs.harvard.edu/abs/2015IJAsB..14..201M} {14, 201}

\bibitem[\protect\citeauthoryear{{Mousis}, {Lunine}, {Tinetti}, {Griffith},
  {Showman}, {Alibert}, {Beaulieu}  \& {Holmes Collaboration}}{{Mousis}
  et~al.}{2009}]{mousis09}
{Mousis} O.,  {Lunine} J.~I.,  {Tinetti} G.,  {Griffith} C.~A.,  {Showman}
  A.~P.,  {Alibert} Y.,  {Beaulieu} J.-P.,   {Holmes Collaboration} 2009,
  \mn@doi [\aap] {10.1051/0004-6361/200913160}, \href
  {http://adsabs.harvard.edu/abs/2009A26A...507.1671M} {507, 1671}

\bibitem[\protect\citeauthoryear{{Mousis}, {Lunine}, {Madhusudhan}  \&
  {Johnson}}{{Mousis} et~al.}{2012}]{mousis12}
{Mousis} O.,  {Lunine} J.~I.,  {Madhusudhan} N.,   {Johnson} T.~V.,  2012,
  \mn@doi [\apjl] {10.1088/2041-8205/751/1/L7}, \href
  {http://adsabs.harvard.edu/abs/2012ApJ...751L...7M} {751, L7}

\bibitem[\protect\citeauthoryear{{Nellis}}{{Nellis}}{1997}]{nellis97}
{Nellis} W.~J.,  1997, Chem.~Eur.~J.~(Germany), Vol.~3, No.~12, p.~1921 - 1924,
  \href {http://adsabs.harvard.edu/abs/1997ChEuJ...3.1921N} {3, 1921}

\bibitem[\protect\citeauthoryear{{Nelson}, {Papaloizou}, {Masset}  \&
  {Kley}}{{Nelson} et~al.}{2000}]{nelson00}
{Nelson} R.~P.,  {Papaloizou} J.~C.~B.,  {Masset} F.,   {Kley} W.,  2000,
  \mn@doi [\mnras] {10.1046/j.1365-8711.2000.03605.x}, \href
  {http://adsabs.harvard.edu/abs/2000MNRAS.318...18N} {318, 18}

\bibitem[\protect\citeauthoryear{{Nettelmann}, {Becker}, {Holst}  \&
  {Redmer}}{{Nettelmann} et~al.}{2012}]{nettelmann12}
{Nettelmann} N.,  {Becker} A.,  {Holst} B.,   {Redmer} R.,  2012, \mn@doi
  [\apj] {10.1088/0004-637X/750/1/52}, \href
  {http://adsabs.harvard.edu/abs/2012ApJ...750...52N} {750, 52}

\bibitem[\protect\citeauthoryear{{{\"O}berg}, {Murray-Clay}  \&
  {Bergin}}{{{\"O}berg} et~al.}{2011}]{oberg11}
{{\"O}berg} K.~I.,  {Murray-Clay} R.,   {Bergin} E.~A.,  2011, \mn@doi [\apjl]
  {10.1088/2041-8205/743/1/L16}, \href
  {http://adsabs.harvard.edu/abs/2011ApJ...743L..16O} {743, L16}

\bibitem[\protect\citeauthoryear{{Opik}}{{Opik}}{1958}]{opik58}
{Opik} E.~J.,  1958, {Physics of meteor flight in the atmosphere.}.
John Wiley \& Sons Inc

\bibitem[\protect\citeauthoryear{{Owen}, {Mahaffy}, {Niemann}, {Atreya},
  {Donahue}, {Bar-Nun}  \& {de Pater}}{{Owen} et~al.}{1999}]{owen99}
{Owen} T.,  {Mahaffy} P.,  {Niemann} H.~B.,  {Atreya} S.,  {Donahue} T.,
  {Bar-Nun} A.,   {de Pater} I.,  1999, \mn@doi [\nat] {10.1038/46232}, \href
  {http://adsabs.harvard.edu/abs/1999Natur.402..269O} {402, 269}

\bibitem[\protect\citeauthoryear{{Papaloizou}, {Nelson}, {Kley}, {Masset}  \&
  {Artymowicz}}{{Papaloizou} et~al.}{2007}]{papaloizou07}
{Papaloizou} J.~C.~B.,  {Nelson} R.~P.,  {Kley} W.,  {Masset} F.~S.,
  {Artymowicz} P.,  2007, Protostars and Planets V, \href
  {http://adsabs.harvard.edu/abs/2007prpl.conf..655P} {pp 655--668}

\bibitem[\protect\citeauthoryear{{Petrovic}, {Markovic}  \&
  {Pavlovic}}{{Petrovic} et~al.}{2003}]{petrovic03}
{Petrovic} S.~T.,  {Markovic} S.,   {Pavlovic} Z.~A.,  2003, \mn@doi [Journal
  of Materials Science] {10.1023/A:1025133904322}, \href
  {http://adsabs.harvard.edu/abs/2003JMatS..38.3263P} {38, 3263}

\bibitem[\protect\citeauthoryear{{Podolak}, {Pollack}  \& {Reynolds}}{{Podolak}
  et~al.}{1988}]{podolak88}
{Podolak} M.,  {Pollack} J.~B.,   {Reynolds} R.~T.,  1988, \mn@doi [\icarus]
  {10.1016/0019-1035(88)90090-5}, \href
  {http://adsabs.harvard.edu/abs/1988Icar...73..163P} {73, 163}

\bibitem[\protect\citeauthoryear{{Pollack}, {Podolak}, {Bodenheimer}  \&
  {Christofferson}}{{Pollack} et~al.}{1986}]{pollack86}
{Pollack} J.~B.,  {Podolak} M.,  {Bodenheimer} P.,   {Christofferson} B.,
  1986, \mn@doi [\icarus] {10.1016/0019-1035(86)90123-5}, \href
  {http://adsabs.harvard.edu/abs/1986Icar...67..409P} {67, 409}

\bibitem[\protect\citeauthoryear{{Pond}, {Palotai}, {Gabriel}, {Korycansky},
  {Harrington}  \& {Rebeli}}{{Pond} et~al.}{2012}]{pond12}
{Pond} J.~W.~T.,  {Palotai} C.,  {Gabriel} T.,  {Korycansky} D.~G.,
  {Harrington} J.,   {Rebeli} N.,  2012, \mn@doi [\apj]
  {10.1088/0004-637X/745/2/113}, \href
  {http://adsabs.harvard.edu/abs/2012ApJ...745..113P} {745, 113}

\bibitem[\protect\citeauthoryear{{Rubinow} \& {Keller}}{{Rubinow} \&
  {Keller}}{1961}]{rubinow61}
{Rubinow} S.~I.,  {Keller} J.~B.,  1961, \mn@doi [Journal of Fluid Mechanics]
  {10.1017/S0022112061000640}, \href
  {http://adsabs.harvard.edu/abs/1961JFM....11..447R} {11, 447}

\bibitem[\protect\citeauthoryear{{Salpeter}}{{Salpeter}}{1973}]{salpeter73}
{Salpeter} E.~E.,  1973, \mn@doi [\apjl] {10.1086/181190}, \href
  {http://adsabs.harvard.edu/abs/1973ApJ...181L..83S} {181, L83}

\bibitem[\protect\citeauthoryear{{Sing} et~al.,}{{Sing} et~al.}{2016}]{sing16}
{Sing} D.~K.,  et~al., 2016, \mn@doi [\nat] {10.1038/nature16068}, \href
  {http://adsabs.harvard.edu/abs/2016Natur.529...59S} {529, 59}

\bibitem[\protect\citeauthoryear{{Svetsov}, {Nemtchinov}  \&
  {Teterev}}{{Svetsov} et~al.}{1995}]{svetsov95}
{Svetsov} V.~V.,  {Nemtchinov} I.~V.,   {Teterev} A.~V.,  1995, \mn@doi
  [\icarus] {10.1006/icar.1995.1116}, \href
  {http://adsabs.harvard.edu/abs/1995Icar..116..131S} {116, 131}

\bibitem[\protect\citeauthoryear{{Todorov}, {Line}, {Pineda}, {Meyer}, {Quanz},
  {Hinkley}  \& {Fortney}}{{Todorov} et~al.}{2015}]{todorov16}
{Todorov} K.~O.,  {Line} M.~R.,  {Pineda} J.~E.,  {Meyer} M.~R.,  {Quanz}
  S.~P.,  {Hinkley} S.,   {Fortney} J.~J.,  2015, preprint, \href
  {http://adsabs.harvard.edu/abs/2015arXiv150400217T} {} (\mn@eprint {arXiv}
  {1504.00217})

\bibitem[\protect\citeauthoryear{{Warner}, {Harris}  \& {Pravec}}{{Warner}
  et~al.}{2009}]{warner09}
{Warner} B.~D.,  {Harris} A.~W.,   {Pravec} P.,  2009, \mn@doi [\icarus]
  {10.1016/j.icarus.2009.02.003}, \href
  {http://adsabs.harvard.edu/abs/2009Icar..202..134W} {202, 134}

\makeatother
\end{thebibliography}


\appendix
\section{Full Angle Mass Ablation Contours}\label{appendixa}
The mass ablation fraction contours for iron and ice planetesimals representative of all impact angles are shown here for completeness. 

\begin{figure*}
\begin{center}
\includegraphics[scale=0.6]{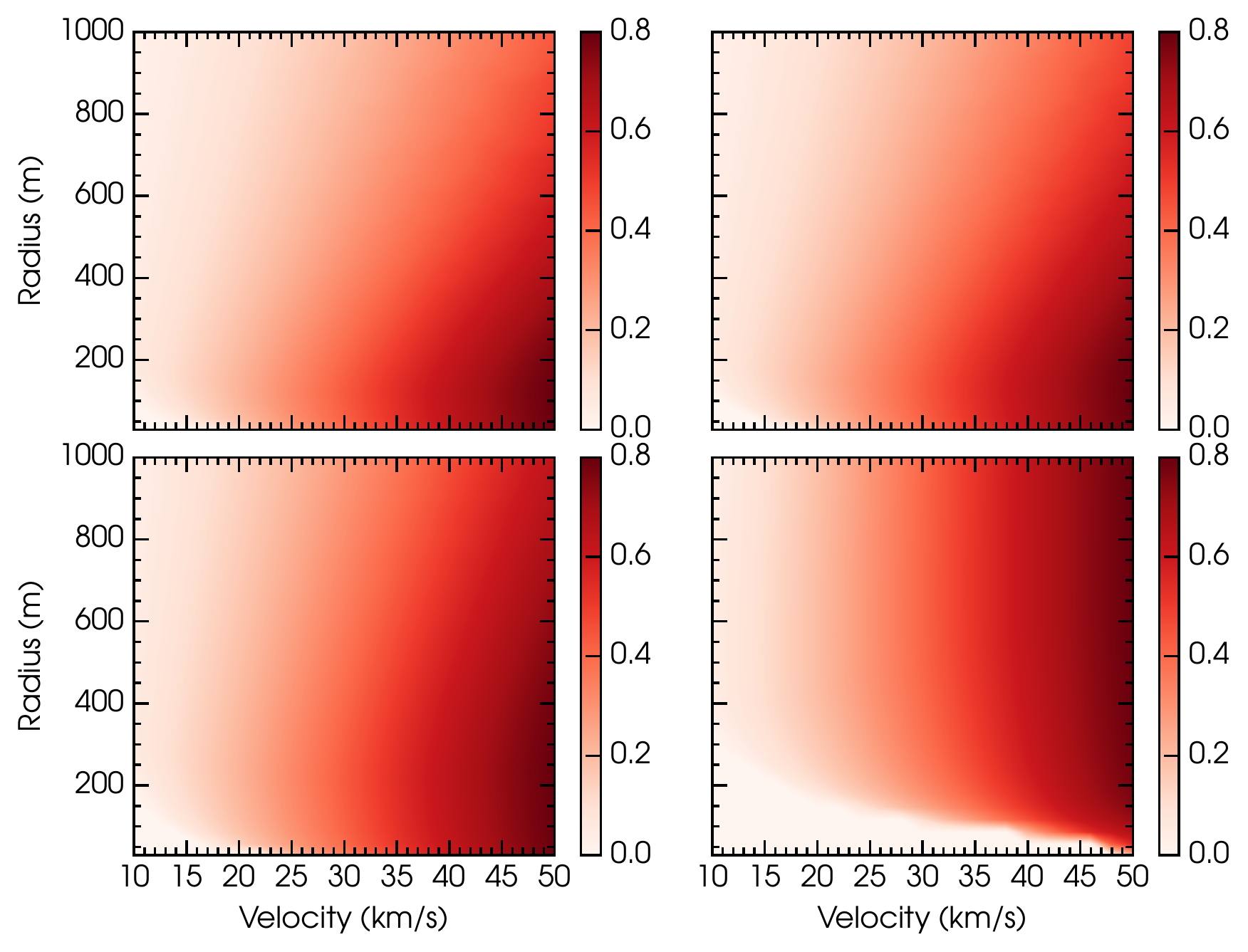}
\caption{Iron planetesimals' mass ablation fraction to a pressure level of 10$^3$ bar for four initial impact angles as a function of initial radius and initial velocity. From top to bottom, left to right: $\phi_i=\{0, 30, 60, 90\}$ degrees from NJP.}

\end{center} 
\end{figure*}    

\begin{figure*}
\begin{center}
\includegraphics[scale=0.6]{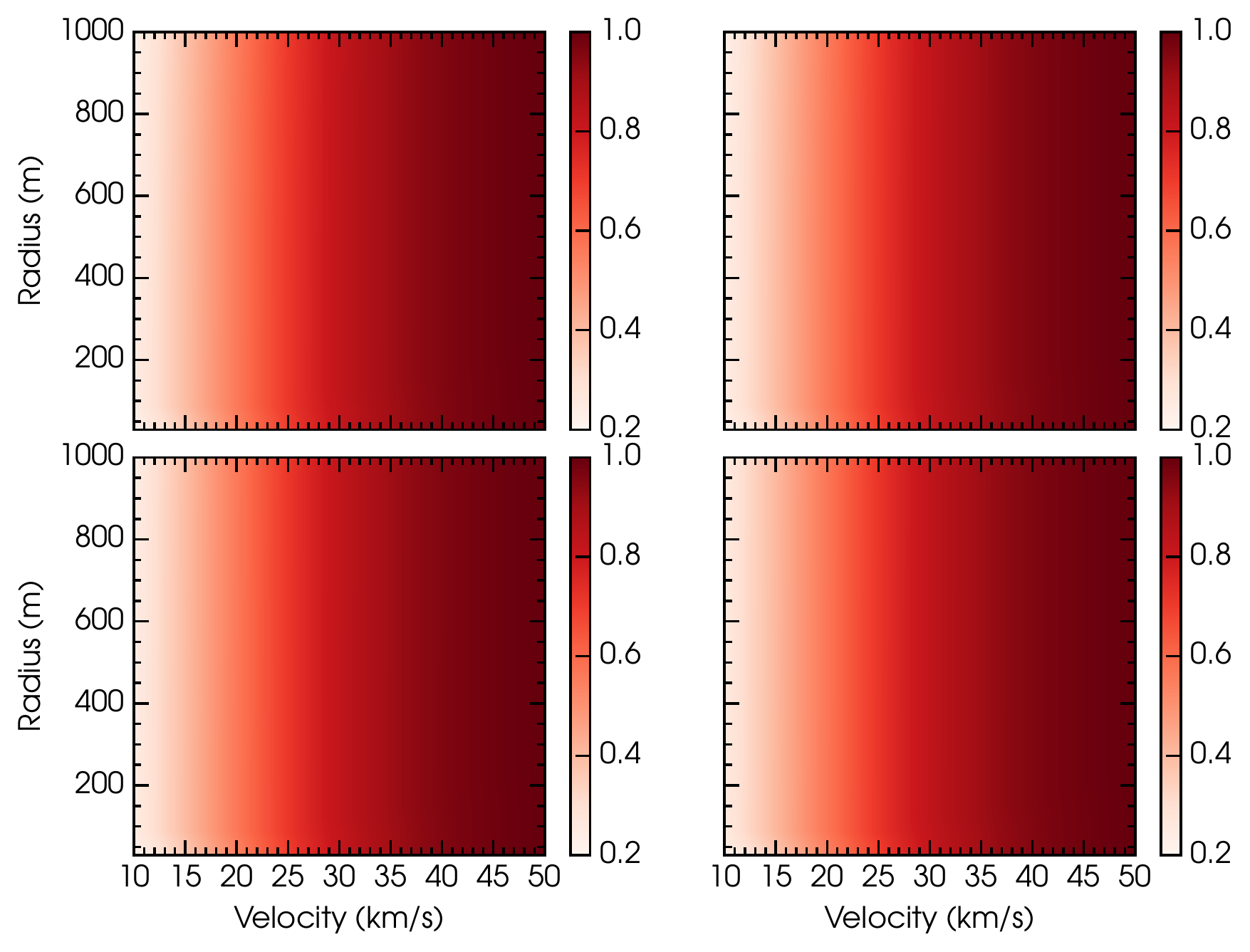}
\caption{Ice planetesimals' mass ablation fraction to a pressure level of 10$^3$ bar for four initial impact angles as a function of initial radius and initial velocity . From top to bottom, left to right: $\phi_i=\{0, 30, 60, 90\}$ degrees from zenith.}

     \end{center} 
\end{figure*}

\begin{figure*}
\begin{center}
\includegraphics[scale=0.6]{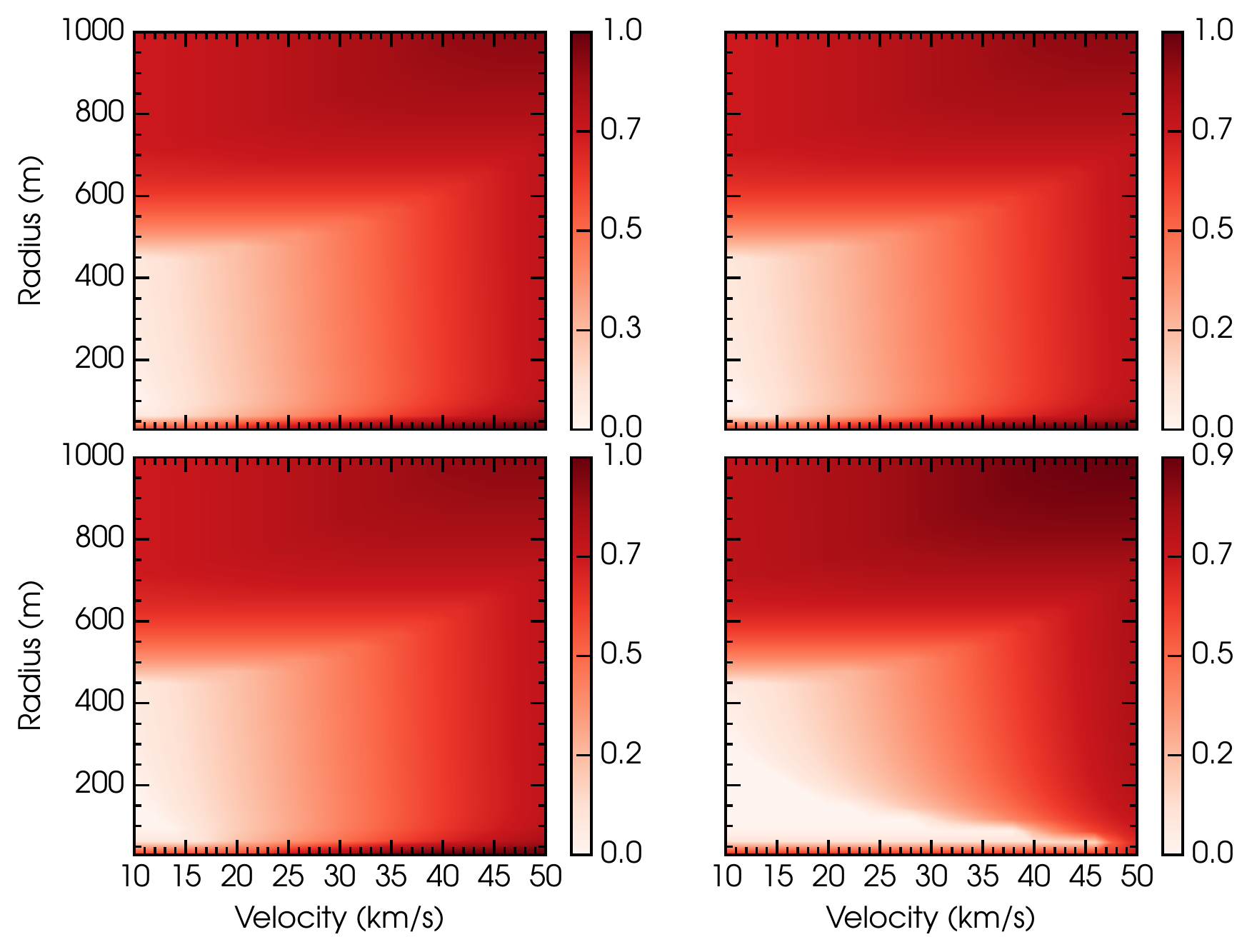}
    \caption{Iron planetesimals' mass ablation fraction to a pressure level of 10 Mbar for four initial impact angles as a function of initial radius and initial velocity. From top to bottom, left to right: $\phi_i=\{0, 30, 60, 90\}$ degrees from zenith.}

     \end{center} 
\end{figure*}
\begin{figure*}
\begin{center}
\includegraphics[scale=0.6]{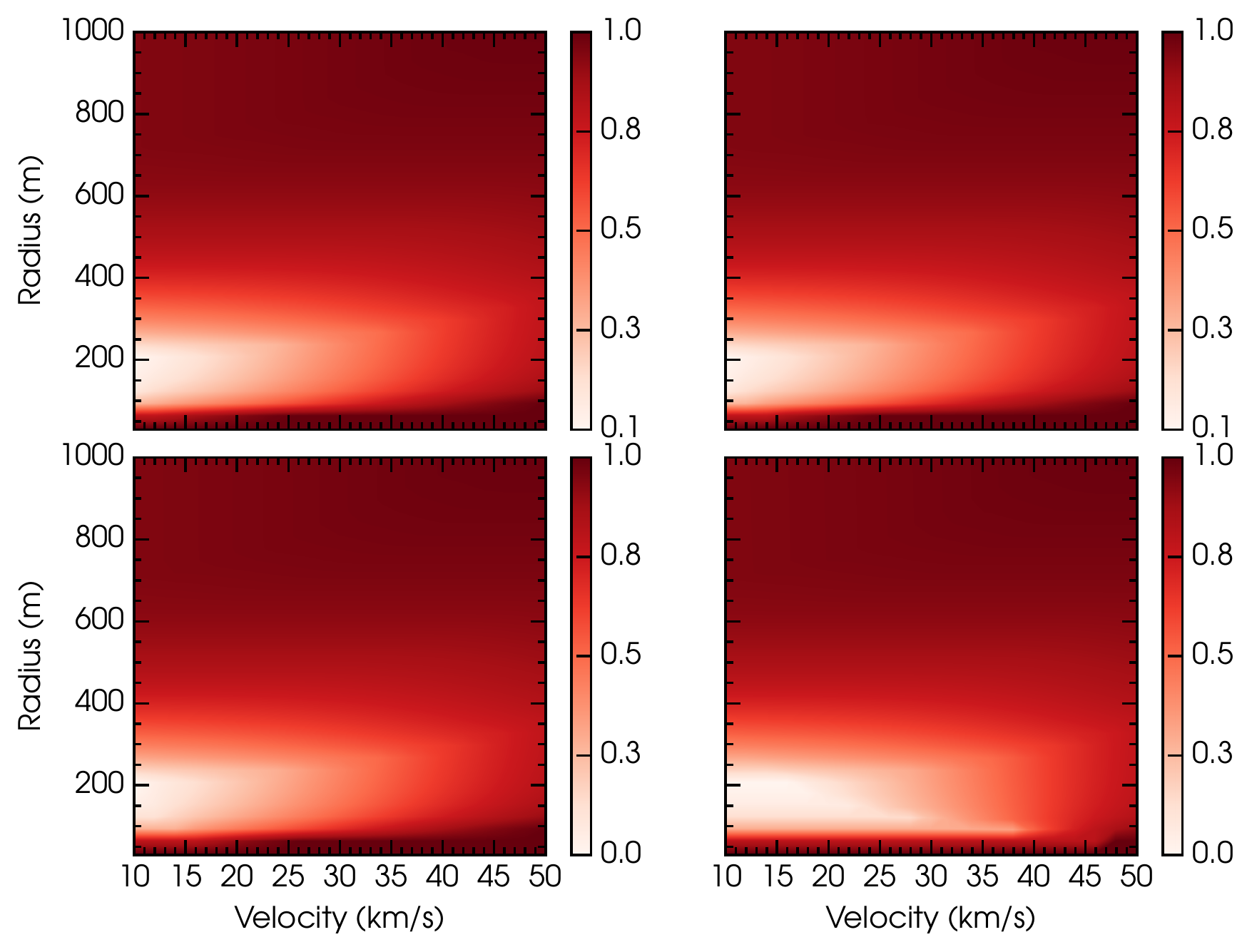}
    \caption{Iron planetesimals' mass ablation fraction to a pressure level coincident with the fiducial core-inner envelope division of 41.687 Mbar for four initial impact angles as a function of initial radius and initial velocity. From top to bottom, left to right: $\phi_i=\{0, 30, 60, 90\}$ degrees from zenith.}

     \end{center} 
\end{figure*}


\bsp	
\label{lastpage}
\end{document}